%% file: main.tex
\documentclass{article}
\usepackage{arxiv}

\usepackage[utf8]{inputenc} % allow utf-8 input
\usepackage[T1]{fontenc}    % use 8-bit T1 fonts
\usepackage{hyperref}       % hyperlinks
\usepackage{url}            % simple URL typesetting
\usepackage{booktabs}       % professional-quality tables
\usepackage{amsfonts}       % blackboard math symbols
\usepackage{nicefrac}       % compact symbols for 1/2, etc.
\usepackage{microtype}      % microtypography
\usepackage{lipsum}		% Can be removed after putting your text content
\usepackage{graphicx}
\usepackage{doi}
\usepackage{amsmath}
\usepackage{xcolor}
\usepackage[normalem]{ulem}
\usepackage{algorithm}
\usepackage{algpseudocode}
\usepackage{braket}  
\usepackage{xcolor}
\usepackage{tikz}
\usepackage[numbers,sort&compress]{natbib}

\usepackage{subcaption}
\usepackage{svg}

\title{Lindbladian Learning with Neural Differential Equations}

\author{
  \normalfont
  Timothy Heightman\thanks{timothy.heightman@icfo.eu}$^{\;\;1,2}$,
  Roman Aseguinolaza Gallo$^{1,3,4}$,
  Edward Jiang$^{1}$, \\
  JRM Saavedra$^{2}$,
  Antonio Ac\'{i}n$^{1,5}$, and
  Marcin P\l{}odzie\'{n}$^{6}$
  \\[6pt]
  $^1$ICFO -- Institut de Ci\`{e}ncies Fot\`{o}niques, The Barcelona Institute of Science and Technology, \\ 08860 Castelldefels, Barcelona, Spain \\
  $^2$Quside Technologies SL, Carrer d'Esteve Terradas 1, 08860 Castelldefels, Barcelona, Spain \\
  $^3$CIC nanoGUNE Consolider, Tolosa Hiribidea 76, E-20018 Donostia-San Sebastian, Spain \\
  $^4$Quantum Motion, 9 Sterling Way, London, N7 9HJ, United Kingdom \\
  $^5$ICREA, Passeig Llu\'{i}s Companys 23, 08010 Barcelona, Spain \\
  $^6$Qilimanjaro Quantum Tech, Carrer de Vene\c{c}uela 74, 08019 Barcelona, Spain
}

\renewcommand{\shorttitle}{}

\hypersetup{
pdftitle={A template for the arxiv style},
pdfsubject={q-bio.NC, q-bio.QM},
pdfauthor={David S.~Hippocampus, Elias D.~Striatum},
pdfkeywords={First keyword, Second keyword, More},
}

\begin{document}
\maketitle

\begin{abstract}
Inferring the dynamical generator of a many-body quantum system from measurement data is essential for the verification, calibration, and control of quantum processors. When the system is open, this task becomes considerably harder than in the purely unitary case, because coherent and dissipative mechanisms can produce similar measurement statistics and long-time data can be insensitive to coherent couplings. Here we tackle this so-called Lindbladian learning problem of open-system characterisation with maximum-likelihood on Pauli measurements at multiple experimentally friendly \emph{transient} times, exploiting the richer information content of transient dynamics. To navigate the resulting non-convex likelihood loss-landscape, we augment the physical model neural differential-equation term, which is progressively removed during training to distil an interpretable Lindbladian solution. Our method reliably learns open-system dynamics across neutral-atom (with 2D connectivity) and superconducting Hamiltonians, as well as the Heisenberg XYZ, and PXP models on a spin-1/2 chain. For the dissipative part, we show robustness over phase noise, thermal noise, and their combination. Our algorithm can robustly infer these dissipative systems over noise-to-signal ratios spanning four orders of magnitude, and system sizes up to $N=6$ qubits with fewer than $5 \times 10^5$ shots. 
\end{abstract}

% keywords can be removed

\section{Introduction}
\label{sec:introduction}

    % \textcolor{purple}{Inferring microscopic properties of a quantum state or process from measurement data alone is inevitably a hard task due to the nondeterministic nature of measurement outcomes, as we all have typicality of quantum states \cite{Bengtsson2006, Haake2010,plodzienHaar2025, cotler2026noisyquantumlearningtheory,schuster2025hardnessrecognizingphasesmatter}. This fundamental obstacle becomes especially concrete in near-term many-body platforms, where one aims to infer device-level dynamics from finite, noisy experimental records.
    % }
    
    % {\bf or :
    % }

Programmable quantum processors realize many-body dynamics whose microscopic parameters---Hamiltonian couplings and dissipative rates---are often only partially known \emph{a priori}. Extracting these device parameters from experimental data is a critical bottleneck in scaling up near-term quantum technologies \cite{Eisert_2020, gebhart2023learning}. Indeed, inferring microscopic properties of a quantum state or process from measurement data is known to be intrinsically difficult \cite{schuster2025hardnessrecognizingphasesmatter,cotler2026noisyquantumlearningtheory}, because measurement outcomes are nondeterministic, and typicality implies that finite data often carry limited information about fine-grained microscopic structure \cite{Bengtsson2006, Haake2010,plodzienHaar2025}.

In \emph{Hamiltonian learning} (HL), one collects finite-statistics measurements on states evolved for known times from known preparations and seeks to infer the unknown Hamiltonian parameters that generate these dynamics. While forward simulation (predicting dynamics given a known Hamiltonian) is straightforward in principle, the inverse map from dynamical measurement data to microscopic parameters is highly nonlinear, and practical inference is constrained by finite data and the exponential growth of Hilbert space. Indeed, brute-force quantum process tomography scales exponentially in parameters and measurements \cite{mohseni2008quantum, Eisert_2020}, motivating approaches that exploit structure such as geometric locality to reduce sample complexity and parameter count. Existing HL methods include equilibrium-based strategies \cite{bairey2019learning, anshu2021sample}, trajectory-based fitting of time-series data \cite{zhang2014quantum, li2020hamiltonian, wilde2022scalablylearningquantummanybody, valenti2022scalable}, and variational schemes based on parametrized circuits \cite{wiebe2014hamiltonian, gupta2023hamiltonian}.

However, realistic quantum hardware is inherently open. Coupling to the environment leads to decoherence and dissipation that are often comparable to coherent timescales in current devices \cite{breuer2002theory, Preskill_2018}. Under standard Markovian assumptions, the dynamics are generated by a time-homogeneous Lindbladian (GKSL generator) rather than a Hamiltonian alone \cite{Lindblad1976, Davies1974, breuer2002theory}. Learning the full Lindbladian constituting coherent couplings and dissipative operators alike, defines the \emph{Lindbladian learning} problem.

Lindbladian learning is substantially more challenging than HL. In addition to Hamiltonian parameters, one must determine dissipative rates (and, in general, channel structure), and identifiability can fail in regimes where distinct generators produce the same, indistinguishable steady-state behaviour. In particular, steady-state data can be largely insensitive to the coherent part of the dynamics, since many Hamiltonians may share the same fixed point \cite{bairey2020learningsteady, Davies1974, Davies1976}; identifiability is therefore a central concern \cite{identifiability2014}. More broadly, the same measurement statistics can often be explained by different coherent--dissipative decompositions, breaking the uniqueness of the inverse map. Existing methods include steady-state approaches \cite{bairey2020learningsteady}, tomography on small processors \cite{Samach_2022}, learning in weakly dissipative regimes \cite{hl_ll_weakly_dissipative_2024}, variational Liouvillian learning \cite{pastori2022characterization}, simulation-assisted strategies \cite{sim_assisted_oqs_2024}, and ansatz-free approaches \cite{ivashkov2026ansatzfreelearninglindbladiandynamics}. However, these methods typically require access to steady states, weak dissipation, or apply only to the smallest system sizes of one or two qubits.

Recent progress to address some of these limitations includes the large-scale time-series protocol of \cite{berg2025large0scale}, which reaches 156-qubit demonstrations by assuming a strictly local Pauli Lindblad model and reducing generator identification to a constrained linear regression on local observables, extending steady-state linearisation of \cite{bairey2020learningsteady} to transient data. As with many hardware-scale identification schemes, this scalability comes with trade-offs. The approach is naturally tied to discrete-depth sampling and relies on extracting dynamical information from fitted time-series, which can reduce sensitivity to fast coherent components outside the resolvable bandwidth and complicate flexible (e.g., non-uniform) sampling. These trade-offs leave open the complementary problem of robust, shot-efficient Lindbladian learning from transient measurement data in regimes where coherent and dissipative contributions induce genuinely non-convex, locally flat or rugged optimization landscapes.

% \tim{Recent work to address these issues includes \cite{berg2025large0scale} which achieve 156-qubit demonstrations by assuming a strictly local Pauli Lindblad model and reducing learning to constrained linear regression on local time-series akin to the linearisation in \cite{bairey2020learningsteady} extended to time-dependent observables. However, this scalability hinges on a tightly structured measurement-and-model pipeline and discrete-depth sampling—whose accuracy is mediated by derivative extraction from curve fits, which is fundamentally bounded by aliasing constraints due to limited flexibility for non-uniform sampling. So coherent terms outside the resolvable bandwidth can be systematically mis-estimated even when data are abundant. Indeed this does not address the fundamentally non-convex nature of Lindbladian Learning.  As such there is a pertinent need for robust Lindbladian Learning algorithms which run on transient data in both weak and strong dissipative regimes which are efficient at the level of shots counts, and require minimal assumptions about quantum optimal control or the convexity of resultant loss landscapes.}

Deep learning offers a promising avenue to overcome these limitations, with applications in quantum science now spanning variational state representation, dynamics prediction, and quantum control \cite{dawid2022modern, heightman2025qml}. In particular, \emph{Neural Differential Equations} (NDEs) provide a principled way to learn continuous-time dynamics by parametrizing the vector field of an ODE with a neural network \cite{chen2018neural, kidger2022nde}. In quantum settings, our previous work \citet{heightman2024solving} showed that augmenting a physics-based Hamiltonian model with an NDE correction can substantially improve optimization and robustness in HL, dynamically reshaping the loss landscape to help variational parameters escape local minima. On open quantum systems \cite{chen2022learning} first proposed NDEs for Lindbladian Learning, although their method requires quantum optimal control, and does not report the shot-intensity or robustness of their method. Neural approaches to quantum dynamics have also been pursued for other aspects of quantum information processing, including latent neural ODEs for learning quantum evolution from partial observations \cite{choi2022learning}, dual-capability models that jointly estimate parameters and predict dynamics \cite{an2025dual}, random-driving protocols combined with deep networks \cite{dl_random_driving_2021}, and machine-learning agents for system characterization \cite{flynn2022quantum}.

In this work we employ NDEs for \emph{Lindbladian learning}: given a known Hamiltonian structure and jump-operator form, we infer all couplings and dissipative rates from finite-shot measurement data. We assume known initial preparations and measurement times, collecting local Pauli snapshots at multiple transient times. We show how maximum-likelihood estimation on this data avoids full process tomography while remaining identifiable under locality assumptions \cite{identifiability2014, shadow2025, Davies1974}. Crucially, transient data is the primary means of resolving the coherent--dissipative ambiguity, since it retains sensitivity to the Hamiltonian component that is lost once the system relaxes toward a fixed point. The NDE then navigates the resulting non-convex optimization landscape. During the augmented training stage the neural correction is constrained to preserve trace and Hermiticity but is not required to maintain complete positivity, so the combined generator can be unphysical during training. However, through a curriculum learning procedure on a combined ansatz and NDE, we are able to yield an output GKSL form generating a strictly CPTP map. Using transient data is also experimentally attractive, as it avoids waiting for convergence to a fixed point, which can be orders of magnitude slower than coherent timescales in low-noise regimes. Unlike \cite{berg2025large0scale}, we resolve the inherent nonconvex optimization pathology that arises when coherent and dissipative mechanisms compete and the loss landscape becomes rugged or locally flat. Further, our method requires no optimal control like \cite{chen2022learning}, and we show robustness over an experimentally friendly maximum of $5\times 10^5$ shots up to $N = 6$ qubits.

To highlight the broad applicability of our protocol, we test it on four physically motivated Hamiltonians. These are the Hamiltonians for neutral-atom (with 2D connectivity) and superconducting QPUs, as well as the Heisenberg XYZ, and PXP models on a spin-1/2 chain. For the dissipative part, we employ three noise models, phase noise, thermal noise, and their combination. We then show how our algorithm can robustly infer these dissipative systems over noise-to-signal ratios spanning four orders of magnitude, and system sizes up to $N=6$ qubits. By systematically comparing physics-only and NDE-augmented models, we identify when neural augmentation improves robustness. We find the NDE-augmentation improving robustness most notably in regimes where non-commuting coherent and dissipative dynamics induce rugged loss landscapes and degenerate fixed-point structure. Conversely, when jump operators commute with the Hamiltonian and the steady state is unique, we find a physics-only model typically suffices and NDEs can overfit. This leads to a practical guideline: attempt physics-only variational learning first when the Hamiltonian and Lindbladian ansatz commute, and deploy NDE augmentation only when robustness is insufficient (which is often when these two components are non-commuting). Furthermore, we show that unlike unitary systems \cite{heightman2024solving}, low state infidelity of open dynamics alone can be misleading. Distinct generators can share the same steady-state manifold, and robustness of parameter recovery is the more informative benchmark. 

The remainder of this paper is organized as follows. In Sec.~\ref{sec:ps}, we formally define the Lindbladian Learning problem in the white-box scenario as an extension of the white-box Hamiltonian learning problem to open systems. In Sec.~\ref{sec:NDEs_open_sys} we describe how to incorporate NDEs into open-system dynamics, detailing our curriculum learning procedure, measurements, and the maximum-likelihood-based loss function. Next, we describe the details of the numerical experiments, datasets, model architectures and benchmarks in Sec.~\ref{sec:numerical_experiments}. Sec.~\ref{sec:results_intro} then presents the results of these numerical experiments, and offers a detailed analysis around the subtleties of Lindbladian Learning through a study of loss-landscape dynamics. Finally, Sec.~\ref{sec:conclusion} offers an outlook and future research directions to scale our algorithm to systems with $N \gg 1$.

\subsection{Problem Statement}
\label{sec:ps}
We consider Markovian open-system dynamics on 1D spin-$1/2$ chains. We assume the system plus environment evolves unitarily and the environment is not observed. Tracing out the environment, the reduced state $\rho(t)$ evolves under a time-homogeneous CPTP semigroup generated by the Lindblad master equation \cite{Lindblad1976},

\begin{equation}
\dot{\rho}(t)=\mathcal{L}_T[\rho(t)]:= -i[H_T,\rho(t)]+\sum_{\alpha}\gamma_\alpha\left(L_\alpha\rho(t)L_\alpha^\dagger-\tfrac{1}{2}\{L_\alpha^\dagger L_\alpha,\rho(t)\}\right),
\label{eq:lindblad}
\end{equation}
with the unitary component generated by Hamiltonian $H_T$, and Lindblad operators ${L_\alpha}$ with non-negative real-valued rates ${\gamma_\alpha}$ generating the open component. We refer to this as the Lindblad master equation, with the right-hand-side commonly referred to as the GKSL form . The learning task is to infer a generator, $\mathcal{L}_T$ in the so-called white-box scenario \cite{heightman2024solving}. This means we set a chosen model ansatz that explains observed data and remains physically valid (CPTP), prioritizing interpretability of both coherent and dissipative terms. The Hamiltonian and dissipative channels are expressed in a fixed, physically motivated operator basis, and the fitted dynamics are attributed to these analytic terms alone. 

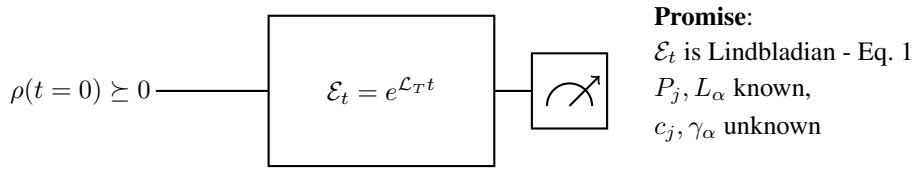
\begin{figure}[h!]
    \centering
\begin{tikzpicture}
    % Draw white box and wires
    \draw[thick] (2,-2.5) rectangle (5,-0.5);
    \node at (3.5,-1.5) {\( \mathcal{E}_t = e^{\mathcal{L}_T t} \)};
    \draw[thick] (0.5,-1.5) -- (2,-1.5);
    \draw[thick] (5,-1.5) -- (5.5,-1.5);
    \node at (-0.5,-1.5) { \(\rho(t=0) \succeq 0\)};
    
    \draw[thick] (5.5,-1.0) rectangle (6.5,-2.0);
    \draw[thick] (6.3,-1.7) arc[start angle=0,end angle=180,radius=0.3];
    \draw[->][thick] (6.0,-1.7) -- (6.4,-1.3);

    % Add Hamiltonian structure known note to the right of the measurement operator
    \node[right] at (7.0,-0.5) {\textbf{Promise}:};
    \node[right] at (7.0,-1.0) { $\mathcal{E}_t$ is Lindbladian - Eq.~\ref{eq:lindblad}};
    \node[right] at (7.0,-1.5) {\(P_j, L_{\alpha}\) known,};
    \node[right] at (7.0,-2.0) {\( c_j,\gamma_{\alpha} \) unknown};
    
\end{tikzpicture}
    \caption{The white-box Lindbladian Learning (LL) setup. The structure of the true Hamiltonian \( H_T = \sum_j c_j P_j \) and open-system terms $\sum_{\alpha}\gamma_\alpha\left(L_\alpha\rho(t)L_\alpha^\dagger-\tfrac{1}{2}\{L_\alpha^\dagger L_\alpha,\rho(t)\}\right)$ structure the ansatz, and the coefficients \( c_j, \gamma_\alpha \in \mathbb{R} \) are unknown.}
    \label{fig:white_box_LL}
\end{figure}

For spin‑1/2 chains of length N, we represent the coherent part in the Pauli basis,
\begin{equation}
H_T=\sum_{j=0}^{4^N-1} c_j P_j,\qquad P_j\in\mathcal{P}_N, c_j\in\mathbb{R},
\label{eq:pauli}
\end{equation}
which is a complete decomposition for any Hamiltonian on N qubits. Here $\mathcal{P}_N$ refers to the Pauli group on N qubits, and dissipative channels are modelled by local jump operators with rates $\gamma_\alpha \in \mathbb{R}_+$. The objective is therefore to learn $({c_j},{\gamma_\alpha})$ so that the induced trajectories match observed statistics within the training window and generalize beyond transient times. As such, our learning algorithm applicable to the white-box scenario as described in \cite{heightman2024solving} with the structure fixed by Lindbladian generator $\mathcal{L}_T$ as shown in Fig.~\ref{fig:white_box_LL}.

We assume only control on the amount of time evolution, and no quantum optimal control. This makes data from time-evolved states experimentally friendly, provided the chosen evolution times are sufficiently short \cite{wilde2022scalablylearningquantummanybody, heightman2024solving}.

\section{Neural Differential Equations in Open Quantum Systems}
\label{sec:NDEs_open_sys}
% {
% This section begins with a description of how to employ NDEs in the context of Lindbladian dynamics. We then proceed to describe a curriculum which removes the dynamics' dependency on the neural component. This makes the output of our learning algorithm interpretable, as we obtain estimators for $\{c_j, \gamma_{\alpha}\}$ that are the only component generating dynamics. Finally this section concludes by detailing how NDEs can be combined with the maximum likelihood method via loss functions and quantum measurements as was done in \cite{heightman2024solving}. 
% \textcolor{purple}{Marcin: I'm not sure we need this paragraph-opener.}

A Neural Ordinary Differential Equation (NDE) specifies the vector field of an ODE with a trainable function. Given a state vector $x(t)\in\mathbb{R}^d$ and a neural architecture $f_\theta:\mathbb{R}^d\to\mathbb{R}^d$ with parameters $\theta$, the NDE dynamics
\begin{equation}
\frac{dx}{dt}(t)=f_\theta\big(x(t)\big),\qquad x(0)=x_0,
\label{eq:NDE}
\end{equation}
are integrated to time $T$, and the parameters $\theta$ are tuned by backpropagating a task loss through the ODE solver. Universal approximation results justify the expressivity of $f_\theta$, and practical training uses stochastic optimizers such as Adam \cite{chen2018neural,kidger2022nde,kingma2014adam}. Despite the GKSL form using partial derivatives, we employ NODEs because when the generator is time-independent, the partial differential equation for $\rho(t)$ becomes an ordinary differential equation. As such an ordinary neural augmentation is sufficient, and we use NDE and NODE interchangeably in the scope of this work.

In our open-system setting the dynamical variable is a density operator $\rho(t)$ evolving under $\mathcal{L}_T$. In line with recent results in Hamiltonian Learning with NDEs \cite{heightman2024solving}, we adopt a gray-box decomposition. Specifically, dynamics are computed by adding two superoperators:
\begin{equation}
\mathcal{L}_{\Theta}=\mathcal{L}_{\mathrm{phys}}(\theta)+\mathcal{L}_{\mathrm{NN}}(\varphi),
\label{eq:liouvillian-sum}
\end{equation}
corresponding to a physics-based ansatz and a neural correction.
The physics ansatz is the standard Lindblad generator,
\begin{equation}
\mathcal{L}_{\mathrm{phys}}(\theta)[\rho]=-i[H(\theta_H),\rho]+\sum_{\alpha}\gamma_\alpha(\theta_\alpha)\Big(L_\alpha\rho L_\alpha^\dagger-\tfrac12\{L_\alpha^\dagger L_\alpha,\rho\}\Big),
\label{eq:lindblad-phys}
\end{equation}
where $\theta_H$ are the Hamiltonian Pauli coefficients, and $\theta_\alpha$ are real-valued free parameters mapped to nonnegative rates by a softplus transformation,
\begin{equation}
\gamma_\alpha(\theta_\alpha)=\mathrm{softplus}\big(\theta_\alpha\big)=\log\big(1+e^{\theta_\alpha}\big).
\label{eq:softplus}
\end{equation}
% This guarantees $\gamma_\alpha\ge 0$ and improves sensitivity of gradients for small rates. 
% \edward{Oh I didn't know we were using softplus here. How are we initializing the $\theta_\alpha$ parameters (more importantly, is this dependent on the noise ratio)?} 
The neural correction $\mathcal{L}_{\mathrm{NN}}(\varphi)$ is implemented as a small, neural-network, acting on $\rho$. An arbitrary neural network does not inherently guarantee that the evolved state remains a valid density matrix. We enforce the physicality of the state during the numerical simulation via explicit projections. To ensure the density matrix maintains unit trace, and is hermitian, we divide the output state by its trace and then average over its conjugate transpose once at the output of the numerical solver (see Fig.~\ref{fig:architecures_all} and Sec.~\ref{sec:neural_arch} for further detail).

% \edward{Same here, why do we do this with the state itself and not the output of the neural term?}.
% \edward{Does this work better than just enforcing that the output of the neural term is traceless? We need to justify this}.

However, explicitly enforcing positive semidefiniteness requires continuous eigenvalue decompositions, which are computationally prohibitive and severely bottleneck the automatic differentiation within the ODE solver. Consequently, we apply no explicit mathematical projection for positive semidefiniteness. Instead, we rely on the empirical observation that, for the physical initial states utilized in our work, the trained network naturally bounds the trajectories, and the density matrices remain positive semidefinite at every point in the simulations. The master equation then reads
\begin{equation}
\dot{\rho}(t)=\big(\mathcal{L}_{\mathrm{phys}}(\theta)+\mathcal{L}_{\mathrm{NN}}(\varphi)\big)[\rho(t)],\qquad \rho(0)=\rho_0,
\label{eq:master-nde}
\end{equation}
and is integrated with an automatic-differentiation-compatible ODE solver. In practice we use a high-order explicit Runge–Kutta integrator with adaptive step control (Diffrax/JAX) \cite{kidger2022nde, rungekutta1996}. The total trainable set of parameters is then $\{\theta, \varphi\}$.

By the universal approximation theorem, this correction term can fully emulate the physics-based ansatz, as well as dynamics not captured by this ansatz. As such, $\mathcal{L}_{NN}$ can dominate the dynamics in training rendering a naive NDE-based learning algorithm uninterpretable at the level of Hamiltonian and Lindbladian coefficients. Hence, it is necessary to heavily control the magnitude of $\mathcal{L}_{NN}$, with an idealised algorithm converging to dynamics whose trajectories are governed only by $\mathcal{L}_{\text{phys}}$ making the output fully interpretable. We can approach this idealised scenario with strong $L_2$ regularization and by a training curriculum described below.

\subsection{Curriculum Learning} \label{sec:curriculum_intro}
The following curriculum gradually reduces reliance on the neural augmentation while refining the interpretable parameters. The rationale behind this is to begin with a more permissive model to navigate rugged landscapes, then constrain capacity to distil an interpretable solution. It works in three steps:
\begin{description}
\item[Warm-up] Jointly train the analytic parameters $\theta$ and the neural parameters $\varphi$. This phase leverages the smoothing effect of the neural term to escape poor basins and accelerate convergence.
\item[Analytic refinement] Switch off the neural term (both forward and backward), reset the optimizer state, and train only $\theta$. The reset is essential to restore sensitivity to small dissipative rates parameterized via the softplus map, ensuring accurate recovery of the interpretable Hamiltonian and Lindbladian coefficients.
\item[Optional residual fine-tuning] With $\theta$ frozen, one may briefly train $\varphi$ to capture small systematic discrepancies not represented in the analytic ansatz. The final weighting of this term can be used as an heuristic indicator of the quality of the ansatz - see Sec.~\ref{sec:conclusion}.
\end{description}

the warm-up removes sensitivity to initialization and barren plateau effects, while the refinement phase consolidates the physically meaningful parameters. We emphasize that we do not impose additional physics-informed penalties beyond the softplus transformation for dissipative rates. Furthermore, in our synthetic studies where the ground truth admits a Lindbladian form we omit this step.

\subsection{Measurements and Loss Function}
\label{sec:NLL_measurements_loss}

As mentioned in the introduction, our learning algorithm is based on maximum-likelihood estimation of learned trajectories, comparing them to measurement data of trajectories evolving under the ground truth dynamics. Here we will discuss in detail how our datasets are constructed as well as the loss function.

To train the NDE, we require a loss function that compares an estimated time-evolved state to observed data in an experimentally friendly manner. Access to ground truth trajectories is therefore assumed only through samples of random local Pauli measurements during transient times, avoiding full process tomography while remaining identifiable under locality assumptions \cite{identifiability2014,shadow2025}. Indeed, transient times speed up the data-gathering process since we don't need to wait for the system to relax into its steady state before measuring. This also means the model gets a good mixed signal from the unitary and dissipative component (because ss data is largely insensitive to the Hamiltonian; many Hamiltonians for the same initial state will collapse to the same (often unique) steady state \cite{Davies1974, Davies1976}.

To that end, let $\rho_\Theta(t)$ denote the model-predicted state at time $t$ obtained by integrating Eq.~\eqref{eq:master-nde} with parameters $\Theta=(\theta,\varphi)$ from a specified initial state $\rho(0)$. For each chosen Pauli measurement basis, we collect binary bitstrings $b\in\{0,1\}^n$, forming a dataset $D$ of measurement outcomes. Following \cite{wilde2022scalablylearningquantummanybody}, with $J$ timestamps, $M$ repeats per measurement, and $K$ random Pauli bases, one has $|D|=MKJ$ samples for a single initial state. Varying the number of initial states multiplies the dataset size correspondingly.

Since our pipeline is numerical, we have full access to $\rho_\Theta(t)$ and can evaluate Born-rule probabilities for observing bitstrings $b$ in the chosen Pauli basis,
\begin{equation}
p_\Theta(b;t)=\mathrm{Tr}\big(\Pi_b\rho_\Theta(t)\big),
\label{eq:born-prob}
\end{equation}
where $\Pi_b$ is the projector associated with $b$ in that Pauli basis. This forms the basis for our loss function, defined as the average negative log-likelihood over the observed outcomes,
% \begin{equation}
% L\big(\rho_\Theta;D\big) = -\sum_{b\in D} \log p_\Theta(b;t),
% \label{eq:nll-simple}
% \end{equation}
\begin{equation}
L\big(\Theta;D\big) = -\sum_{(\rho_0,t,\mathcal{P}_N,b)\in D} \log p_\Theta(b;\rho_0,t,\mathcal{P}_N),
\label{eq:nll-simple}
\end{equation}
where $D$ is a dataset of outcomes acquired from evolving the corresponding a set of initial states $\rho_0$ up to time $t$ with Pauli measurement $P$ with binary outcome $b$ (see Sec.~\ref{sec:data} for further detail on datasets). This loss is differentiable with respect to $\Theta$ through the ODE solver, enabling end-to-end gradient-based updates of the analytic parameters $\theta$ and the neural parameters $\varphi$ \cite{kidger2022nde}.

% \begin{algorithm}[h!]
% \caption{NDE Lindbladian Learning Train Step}\label{alg:LL_NDEs}
%     \begin{algorithmic}[1]
%         \State \textbf{Input} $\mathcal{D}$, $\mathcal{L}_{\text{phys}}$,  $\mathcal{L}_{\text{NN}}$\Comment{Variational parameters $\Theta = \{\theta, \varphi\}$, $\mathcal{D}$ is batched data }
%         \For{Each initial state $\rho(0)$ in $\mathcal{D}:$} 
%             \State Numerically integrate 
%             $\rho(0)$ to $\tilde{\rho}_{\Theta}(t)$
%             \State Compute Loss $ L({\tilde{\rho}_{\Theta}(t)}; \mathcal{D})$
%             \State $\Theta \gets \Theta - \nabla_{\Theta} L({\tilde{\rho}_{\Theta}(t)}; \mathcal{D})$ \Comment{Any optimisation routine can work here}
%         \EndFor
%     \end{algorithmic} 
% \end{algorithm} 

\section{Numerical Experiments and Results}
\label{sec:numerical_experiments}
In our numerical experiments, we will compare a NDE model to a base (vanilla) model. The vanilla model contains only variational parameters, $\theta$, which are coefficients of operators in the GKSL form in Eq.~\ref{eq:lindblad}, whereas the NDE model contains both GKSL variational parameters $\theta$ and neural parameters $\varphi$. The differences between these two architectures are shown in Fig.~\ref{fig:architecures_all}.

This section specifies the dynamical models and the data protocol used throughout. We consider a broad class of systems, with three types of noise whose strength vary over four orders of magnitude. We begin by laying out the Lindbladian terms considered, then proceeding to the four Hamiltonian models. We then detail how datasets are generated, our choice of neural architecture and training hyperparameters, and finally lay out the benchmarks we will use to assess performance. 

\subsection{Dynamical Models and Test Cases}
\label{sec:test_cases}

The two types of noise considered are phase damping, where $\mathcal{L} = \sigma_{z}$, and thermal noise. That is, we couple each site to a thermal-like environment using local ladder jump operators defined as
\begin{equation}
L_i^-=\tfrac{1}{2}(X_i-iY_i),\qquad L_i^+=\tfrac{1}{2}(X_i+iY_i),
\end{equation}
with nonnegative rates $\gamma_i^-,\gamma_i^+\ge 0$, implementing amplitude damping and thermal excitation. For Lindbladian coefficients, we sample true values $\gamma \sim \mathcal{U}(0.2, 1)$. We will also test \textit{combined} thermal and phase noise, where both disspative terms are enabled.

On the Hamiltonian side, we opt for two models of experimental relevance in the quantum hardware community. These are the  Rydberg Hamiltonian from neutral atom QPUs \cite{silverio2022pulser},
\begin{equation}
    H_{\text{R}} = \tfrac{1}{2} \, \Omega \sum_j X_j 
- \delta \sum_j Z_j 
+ \sum_{i \neq j} \frac{C_6}{r_{ij}^6} Z_i Z_j.
\end{equation}
and qubit Hamiltonian for superconducting circuits 
\begin{equation}
    H_{\text{SC}} = \sum_i \tfrac{1}{2} h_i (\mathbb{I}_i - Z_i) + \sum_{\langle i,j \rangle} \tfrac{1}{2}\zeta_{ij}Z_i Z_j 
\end{equation}
This means that the Hamiltonian component has variational parameters $\theta = \{\Omega,\delta, C_6\}$ for the Rydberg model, and $\theta = \{h_i, \}_{i=1}^N \cup \{\zeta_{ij}\}_{\braket{i,j}}$ for the superconducting model.

Indeed combined with the two Lindbladian terms above, the full model captures finite-temperature re-excitation processes and dominant $T_1$-type relaxation respectively in neutral atom  and superconducting QPUs \cite{quantum_engineer_sc_review, Liu2025}. This is because there are variational parameters on both the Hamiltonian component ($\theta$) and on the dissipative part ($\gamma$). We note here that thermal channels also induce unique steady states that regularize long-time dynamics \cite{Davies1974, Davies1976}, improving identifiability of both coherent and dissipative parameters from short-time transients and stabilizing extrapolation beyond the training window. These properties align with steady-state identifiability results and transient-time learning strategies reported in recent open-system learning literature \cite{hl_ll_weakly_dissipative_2024, chen2025transformer}. 

For the neutral atom TFIM, the true parameters to-be-learned are sampled uniformly in ranges applicable to neutral atom QPUs \cite{silverio2022pulser}, $\Omega \sim \mathcal{U}(0, 1)$, $\delta \sim \mathcal{U}(-4, 4)$, with $C_6 = 10^6\,\mu\text{m}^6$. Atoms are arranged into a compact triangular (hexagonal) lattice with spacing $a \sim \mathcal{U}[9, 11] \, \mu\text{m}$. Positions are perturbed by $\pm 5\%$ of $a$, so distances differ slightly from the ideal lattice and individual interactions have to be learnt. We choose to model nearest-neighbour and next-nearest-neighbour interactions. 

For the superconducting Hamiltonian, true parameters are sampled from normal distributions $h/2\pi \sim \mathcal{N}(0, 10)\,\text{kHz}$ and $\zeta/2\pi \sim \mathcal{N}(-30, 10)\,\text{kHz}$. Values are scaled by a factor of $1/100$ to normalize the energy scale to $\mathcal{O}(1)$, meaning our results are applicable to current superconducting QPUs \cite{schwartzman2025modeling, abughanem2025superconducting, huang2020superconducting}.

Finally, to stress-test our algorithm, we also study spin-1/2 chain Hamiltonians. Following previous benchmarks for challenging unitary dynamics in this context \cite{heightman2024solving}, we select two models that probe complementary regimes of identifiability and dynamical complexity: (i) the anisotropic Heisenberg chain,
\begin{equation}
H_{\mathrm{H}}=\sum_{i=1}^{N-1}\big(J_i^x X_iX_{i+1}+J_i^y Y_iY_{i+1}+J_i^z Z_iZ_{i+1}\big)+\sum_{i=1}^{N} h_i X_i,
\end{equation}
and (ii) the PXP model \cite{Serbyn2021}, a paradigmatic weakly non-ergodic Hamiltonian,
\begin{equation}
H_{\mathrm{PXP}}=\sum_{i=2}^{N-1} J_i P_{i-1} X_i P_{i+1},\quad \text{with}\quad P_i=|{0}\rangle \langle0|_i.
\end{equation}
Both were previously identified as hard instances for closed-system HL due to rugged loss landscapes and weakly non-ergodic dynamics, respectively, serving as stringent tests for robustness and generalization of our open-system learning algorithm \cite{heightman2024solving}. True values for these two models are sampled uniformly, for Hamiltonian coefficients $J_i \sim \mathcal{U}(-1, 1) \; \forall i$. To further stress-test our algorithm, we split the unitary and noise terms of Eq.~\ref{eq:lindblad}, and define a multiplier $R$ for the ground truth values sampled uniformly. After sampling true parameters, we rescale them such that the multiplier induces a noise-to-unitary ratio $R \in \{0.01, 0.1, 1, 10\}$, allowing us to test over four orders of magnitude that interpolate between weak and strong dissipation.

\subsection{Datasets}
\label{sec:data}
Ground-truth datasets are generated by numerically integrating the Lindblad master equation with the specified $H$ and noise term, using an adaptive high-order explicit integrator applied directly to the density matrix representation. Unless stated otherwise, Hamiltonian parameters are drawn i.i.d. from $[-1,1]$, and dissipative rates are set by sampling $\gamma_i^-,\gamma_i^+$ in $[0.2,1.0]$ and rescaling by a global noise-to-unitary level $R \in \{10^{-2}, 10^{-1}, 10^0, 10^1\}$ to probe robustness across weak-to-strong dissipation. For each $N\in\{3,4,5,6\}$ we prepare $L=5$ product eigenstates of single-qubit Pauli operators, drawn uniformly from the sets of $X$-, $Y$- and $Z$-eigenstates (i.e., “up”/“down” along $x,y,z$), and evolve to $J=10$ timestamps $t\in\{0.1,0.2,\dots,1.0\}$ (in units where $\hbar=1$). Measurements follow the protocol from \cite{heightman2024solving}: at each $(\text{state},t)$ pair, we choose $K=200$ random compatible Pauli bases and collect $M=100$ shots per basis. This means each numerical experiment uses no more than $5 \times 10^5$ shots in total. We test every combination of Hamiltonian and noise, for each value of $\eta$ meaning 40 different numerical experiments for each value of $N$, and a total of 200 experiments when $3 \leq  N \leq 6$. For robustness, we generate $50$ independent ground-truth parameter sets $\theta$ for each of the 200 experiments.

Training uses the average negative log-likelihood under the Born rule computed from predicted $\rho(t)$ and the observed bitstrings per Sec.~\ref{sec:NLL_measurements_loss}. This transient, random-Pauli scheme is consistent with identifiability from local measurements \cite{berg2025large0scale, bairey2020learningsteady}, avoiding oracle access, optimal control \cite{chen2022learning}, and steady-state requirements such as those in \cite{bairey2019learning} .\\

\subsection{Neural Architecture}
\label{sec:neural_arch}

When used, the NDE component is a simple, shallow Multi-Layer-Perceptron (MLP) with bias \cite{dawid2022modern, heightman2025deep}. As mentioned in Sec.~\ref{sec:NDEs_open_sys}, we employ a scaled soft-plus activation, $f(x) = \frac{1}{5}\log(1+e^{5x})$ for the neural network component. We initialise the variational parameters for the Hamiltonian  by sampling $\theta \sim \mathcal{U}(-1, 1)$, where $\theta$ are the variational parameters of a given Hamiltonian model from Sec.~\ref{sec:test_cases}. Meanwhile for the dissipative component we initialise $\gamma \sim \mathcal{U}(0.2, 1)$. This means the structure of the Lindbladian is fixed and only the variational coefficients are fitted in line with the white-box scenario \cite{heightman2024solving}.

To train, we integrated the Lindblad master equation using a 5th-order explicit Runge-Kutta method 
\cite{tsitouras2011runge} using the 
diffrax package \cite{kidger2022nde, kidger2021on} from the JAX ecosystem \cite{jax2018github}. We utilize the Adam optimizer for all training phases \cite{kingma2014adam}. 

Each training epoch consists of 500 optimization steps. The dataset is randomly shuffled at the start of every epoch. At each step, the model processes a batch consisting of  1 initial state (out of 5), 1 single measurement shot (out of 100), all 10 time points, and all 200 Pauli observables. This yields $5 \text{ states} \times 100 \text{ shots} = 500$ steps per epoch. Per Sec.~\ref{sec:curriculum_intro} we split training into three phases. Finally, we detail the curriculum learning strategy for both the vanilla and NDE component and their respective learning rates $\eta$..

\begin{itemize}
    \item \textbf{Vanilla Training Curriculum:} 
    \begin{enumerate}
        \item 20 epochs with learning rate $\eta_H = \eta_L = 10^{-3}$. 
        \item \textit{Fine-tuning (if required):} Optimizer reset, followed by 5 epochs with $\eta_H=10^{-4}, \eta_L=10^{-3}$, and a final reset for 5 epochs with $\eta_H=\eta_L=10^{-4}$. (Rates are swapped if the Lindbladian magnitude exceeds the Hamiltonian).
    \end{enumerate}
    \item \textbf{NDE Training Curriculum:}
    \begin{enumerate}
        \item 20 epochs with $\eta_H = \eta_L = 10^{-3}$, $\eta_{\text{NDE}} = 2\times 10^{-3}$, and L2 regularization $\lambda = 0.1$.
        \item 10 epochs with the NDE switched \textbf{OFF}, continuing with $\eta_H = \eta_L = 10^{-3}$ after an optimizer reset.
        \item \textit{Fine-tuning:} Identical to the Base Training regime.
    \end{enumerate}
\end{itemize}

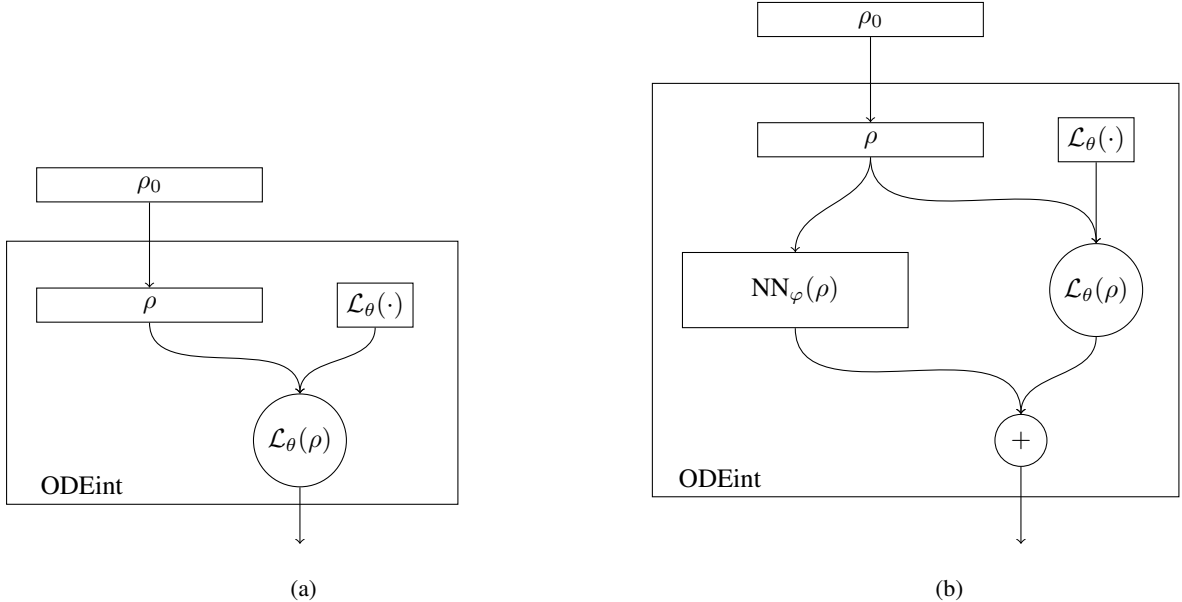
\begin{figure}[h!]
    \begin{subfigure}{0.48 \textwidth}
      \input{Figures/architectures/vanilla.tikz}
      \caption{}
      \label{fig:architectures_a}
    \end{subfigure}
    \hfill
    \begin{subfigure}{0.48 \textwidth}
      \input{Figures/architectures/NDE.tikz}
      \caption{}
      \label{fig:architectures_b}
    \end{subfigure}
     \caption{Computational graphs for automatic differentiation of an ordinary differential equation solver \texttt{ODEint} for a vanilla model (a), and a NDE model, (b). Here by ODE-int we refer to using a numerical solver compatible with automatic differentiation. The subscripts $\theta$ and $\varphi$ indicate the variational parameters assigned to the Lindbladian and Neural component respectively in both (a) and (b). As mentioned in Sec.~\ref{sec:NDEs_open_sys}, we enforce physical output states by renormalising via the trace and averaging over the conjugate transpose once on the output of the ODEint numerical solver.}
     \label{fig:architecures_all}
\end{figure}

Crucially, the optimizer state is reset whenever learning rates are updated or when the NDE component is toggled to allow gradient updates to adapt freshly to the new loss landscape. The architectures for vanilla and NDE models as computational graphs for automatic differentiation are shown in Fig.~\ref{fig:architecures_all}.

\subsection{Performance Metrics}
In order to test the effectiveness of the vanilla and NDE models, we realise 50 numerical experiments for each combination of Hamiltonian and dissipative components. In each experiment, we resample the true-parameters and re-initialise the variational parameters to explore how robust each architecture is over these different realisations. Following \cite{heightman2024solving} we make a scalar robustness metric as follows: Let $\tilde{\theta}$ denote the learned estimator after training via the chosen learning algorithm (vanilla or NDE-enhanced). We calculate the relative absolute error for the coherent parameters,
\begin{equation}
\epsilon_{H}(\tilde{\theta})=\frac{| \theta_H-\tilde{\theta}_H|_1}{|\theta_H|_1},
\end{equation}
and, when Lindbladian parameters are learned, the analogous error for the dissipative parameters,
\begin{equation}
\epsilon_{L}(\tilde{\theta})=\frac{| \theta_L-\tilde{\theta}_L|_1}{|\theta_L|_1}.
\end{equation}
A trial is declared successful for parameters $S\in{H,L}$ if $\epsilon_S(\tilde{\theta})<0.1$. This threshold mirrors prior practice and cleanly separates converged from failed runs across all experiments. Reported success rates are the proportion of successful trials over the $50$ draws for fixed $(H,N)$ and fixed experimental condition (e.g., noise level, learning algorithm).\\

To quantify generalization, we evaluate the density-matrix fidelity 
\begin{equation}
F(\rho_{\mathrm{T}}(t),\rho(t;\theta))=\left(\mathrm{Tr}\sqrt{\sqrt{\rho_{\mathrm{T}}(t)}\rho(t;\theta)\sqrt{\rho_{\mathrm{T}}(t)}}\right)^2
\end{equation}
and define the infidelity loss
\begin{equation}
\mathcal{I}\big(\rho_{\mathrm{T}}(t),\rho(t;\theta)\big)=1-F\big(\rho_{\mathrm{T}}(t),\rho(t;\theta)\big).
\end{equation}

\section{Results}
\label{sec:results_intro}

% In this section, we show the results for the systems and noise types outlined in Sec.~\ref{sec:numerical_experiments}. We begin by discussing trennds and results that apply to all systems, then showing each system's results and benchmarks in Secs.~\ref{sec:NA_results}-\ref{sec:PXP_results}. In each case, we will show the robustness benchmark for the Hamiltonian and Lindbladian parameters separately, as well as showing how each system's robustness is affected by the choice of Lindbladian structure and amplitude by analysing loss landscapes. We then show the infidelity benchmarks.\textcolor{purple}{Marcin: not shure we need paragraph opener}
 
Over the orders of magnitude of noise, $R \in \{0.01, 0.1, 1, 10\}$, the robustness is highest for systems with high noise in both the vanilla and NDE architectures. This is because the system thermalises more rapidly to a steady state. Hence, the transient evolution data used for inference gives a larger representative window of the dynamics over the snapshot times. As such, we found the vanilla model to be sufficiently robust, sometimes converging on $100\%$ of our tests (see e.g. Figs~\ref{fig:NA_TFIM_robustness} and \ref{fig:SC_robustness}) for $R = 1, 10$, especially in phase and thermal noise. However, over smaller amounts of noise $R = 0.01, \;0.1$, we see the NDE architecture significantly outperforming the vanilla architecture on phase and thermal noise, with the performance gain typically increasing as noise was decreased (see e.g. Fig.~\ref{fig:NA_TFIM_robustness}). In line with \cite{heightman2024solving}, we can see what NDEs are providing to this learning problem through loss landscapes. Unlike the unitary dynamics' landscapes in \cite{heightman2024solving}, we found that the dissipative dynamics' loss-landscapes were initially rugged, even when the NDE was improving performance. 

To that end, it is instructive to define the variational space 
\begin{equation}
    V = \mathbb{R}^{\dim(c_j)} \times \mathbb{R}^{\dim(\gamma_{\alpha})} \times \mathbb{R}^{\dim(\varphi)},
    \label{eq:variational_subspaces}
\end{equation}
consisting of three subspaces that respectively correspond to Hamiltonian ($c_j$), Lindbladian ($\gamma_{\alpha}$) and neural variational parameters ($\varphi$). This means the vanilla model is variational over only $\mathbb{R}^{\dim(c_j)} \times \mathbb{R}^{\dim(\gamma_{\alpha})} \subset V$ and the NODE architecture is variational over all of $V$ via the different stages of the curriculum defined in Sec.~\ref{sec:neural_arch}. Having defined $V$, we can construct random orthogonal directions, $v_1,v_2 \in V$, as well as random orthogonal directions to explore how the variational landscape behaves for the Hamiltonian (we refer to these as $x_1,x_2 \in \mathbb{R}^{\dim(c_j)}$, Lindbladian $y_1,y_2\in \mathbb{R}^{\dim(\gamma_{\alpha})}$, and neural parameters $\varphi_1, \varphi_2 \in \mathbb{R}^{\dim(\varphi)}$, and any combination thereof, such as $z_1,z_2 \in \mathbb{R}^{\dim(c_j)} \times \mathbb{R}^{\dim(\gamma_{\alpha})}$. 

Indeed when we plot random orthogonal directions in this subspaces of $V$, we see a richer dynamical structure to the way in which neural methods are offering gains. We found the NDE loss-landscapes to smoothen the Hamiltonian and dissipative loss landscapes later into training epochs, rather than over entire training runs. See e.g. Fig.~\ref{fig:XYZ_loss_in_epochs}.

For the case of both noise types, performance of the vanilla model degrades slightly on the Lindbladian parameters, and significantly on the Hamiltonian parameters. We also found the NDE performance to be generally poorer in this regime. This is because when many types of noise are present in the dynamics, the measurement data stops being sensitive to variations in the hamiltonian parameters when the system acquires more and/or stronger dissipative variational parameters. We see this in the loss-landscapes of the Hamiltonian parameters in this noise scenario, which show a barren landscape around the true value which is at the origin of each plot (see e.g. Fig.~\ref{fig:NA_TFIM_phs_H_loss_landscape}). Likewise, the NDE correction cannot restore sensitivity of the Hamiltonian parameters in this case because this insensitivity is in the training data itself. 

The advantage of strong noise combined with transient data is especially visible in the infidelity benchmarks, where systems showed better infidelities for higher amounts of noise. Even when we renormalise the training window's snapshot times to the noise magnitude, we see systems with stronger noise showing better infidelities (see e.g. Fig.~\ref{fig:NA_infidelities_all}). This is also because, unlike unitary evolution, the infidelity stabilises much lower when the learning algorithm converges to the correct steady state. Indeed, for many system considered, the steady state is unique due on the commutant structure of the Hamiltonian and dissipative component of the Lindbladian \cite{Davies1974, Davies1976}. When this is the case, the ODE model infidelities of flatten out, and we see the systems reaching their steady states at the same point on the renormalised infidelity points in Fig.~\ref{fig:NA_infidelities_all}. In fact, the similar profile of each systems infidelity curves indicates model's error keeps increasing until the system relaxes to its fixed point and many variational parameters become irrelevant. Indeed this further justification for our use transient times for measurements constructing the training dataset from a learning perspective on top of being experimentally-friendly. If we used only measurements based on steady-state data, the model would only need to correctly learn how to reach the steady state, not \textit{which} transient path was taken, as we highlighted in Sec.~\ref{sec:NLL_measurements_loss}. We see some non-zero infidelity because the learned model's steady-state and true steady-state still differ, usually by $<0.1\%$ for many of the cases that follow.

We will now proceed to a more detailed study of each system's benchmarks.

\subsection{Neutral Atom (TFIM) Hamiltonian}
\label{sec:NA_results}
\begin{figure}
    \centering
    \includegraphics[width=1\linewidth]{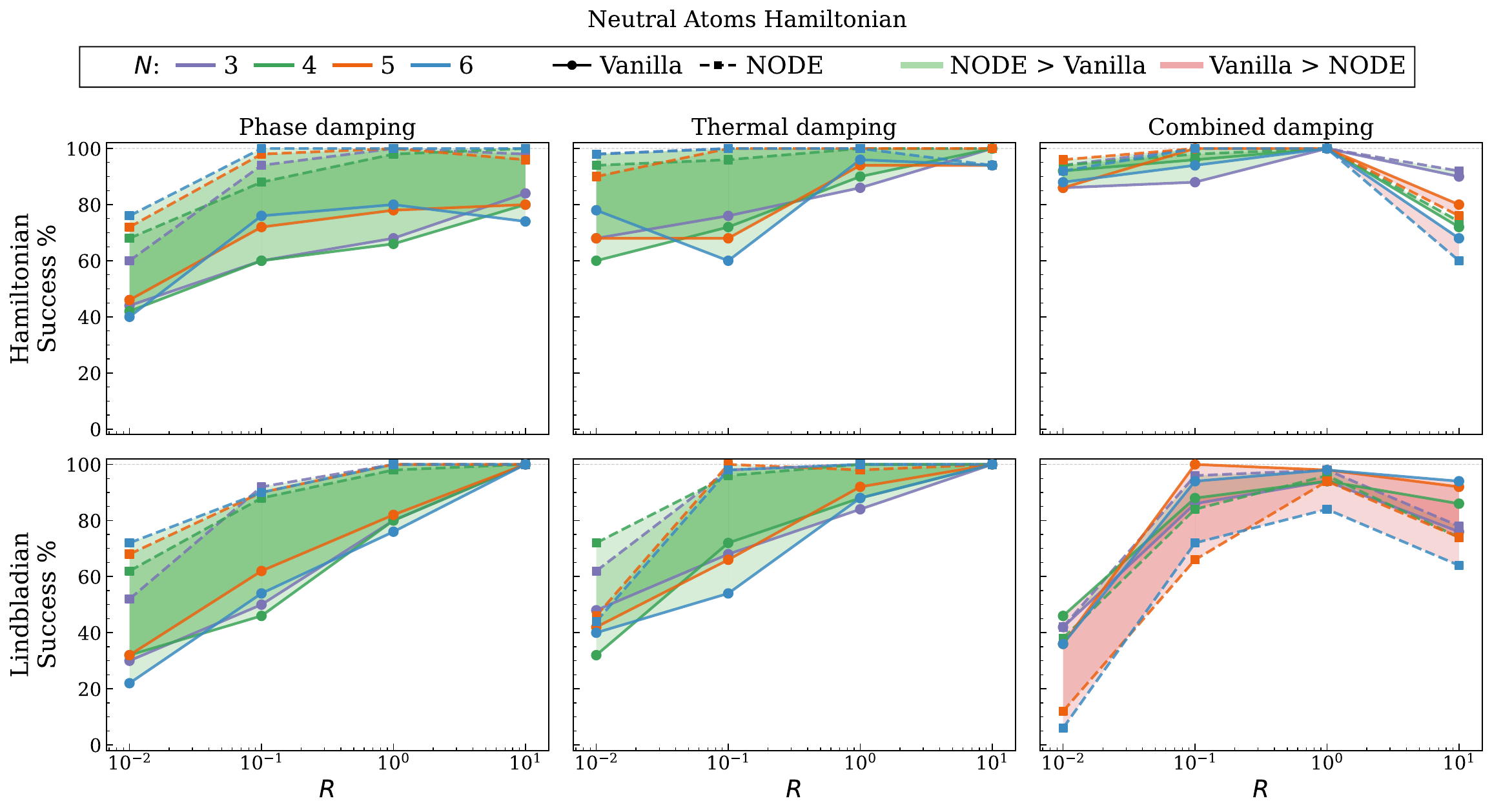}
    \caption{Success rates for a 1D Transverse-Field Ising Hamiltonian model with phase (left), thermal (centre) and combined phase and thermal (right) noise terms over four orders of magnitude of noise-to-unitary ratios, $R$. The top row shows the robustness metric over the Hamiltonian parameters and the shows the robustness metric for Lindbladian parameters. Shaded regions correspond to when the NDE term increases (red) or decreases (green) the fitting error.}
    \label{fig:NA_TFIM_robustness}
\end{figure}

Here, Fig.~\ref{fig:NA_TFIM_robustness} shows the NDE architecture improves success rates considerably for phase and thermal noise. We see this improvement in prediction accuracy on both  the Hamiltonian and dissipative variational parameters, with the gap in performance between the vanilla and NDE model widening as the ratio between the Hamiltonian and dissipative part decreases to the minimum of $10^{-2}$ considered in this work. Indeed at this noise level, the vanilla architecture failed over all 50 test cases. Whereas the NDE architecture saw significantly improved robustness, with its best change in performance being on thermal noise. Curiously, Fig.~\ref{fig:NA_TFIM_robustness} shows that the robustness of the NDE model is \textit{increasing} with $N$, as shown by the $N = 6$ (yellow-square) case sitting firmly above its $ N < 6$ counterparts for all variational components except the thermal dissipative component (Fig.~\ref{fig:NA_TFIM_robustness} lower central). This result, while unexpected, could be a finite-size effect, or a property of thermal dissipative dynamics. Due to computational restrictions, we were unable to run these experiments past $N = 6$, as for $N = 6$ the numerical integration scheme is already running over $4^6 = 4096$-dimensional vectors. See Sec.~\ref{sec:conclusion} for further discussion on scaling.

Next, we turn our attention to the combined noise case showing in Fig~\ref{fig:NA_TFIM_robustness}, in which the Hamiltonian component saw little improvement with NDEs and the dissipative component was actually worsened. This indicates that there are some regimes in which an NDE can improve the inference prospects of a Lindbladian Learning (LL) problem, and others where it becomes worse. Indeed, following the analysis of \cite{heightman2024solving} we can get a more detailed view of when an NDE improves or worsens a LL problem by examining the loss landscapes of the models considered.

\begin{figure}[htbp]
     \centering
     \begin{subfigure}[b]{0.49\textwidth}
         \centering
         \includegraphics[width=\textwidth]{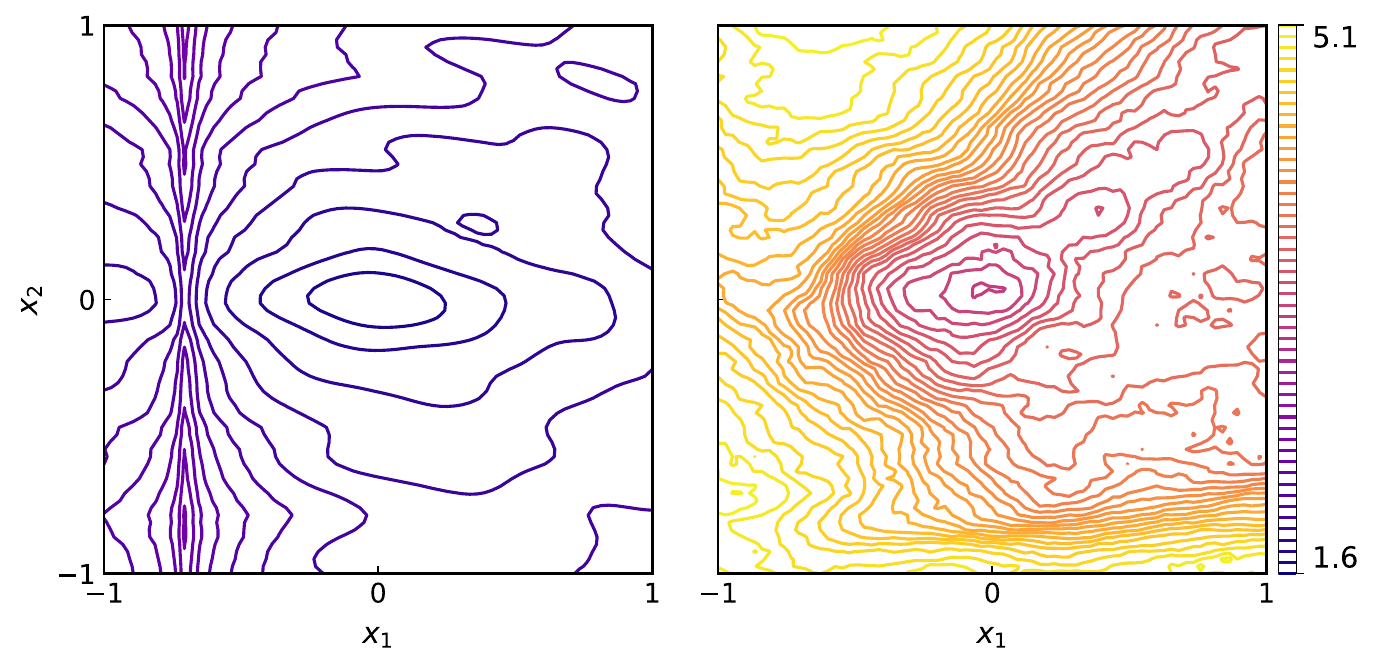}
         \caption{$R = 0.01$ noise} % ($t=10$)
     \end{subfigure}
     \hfill
     \begin{subfigure}[b]{0.49\textwidth}
         \centering
         \includegraphics[width=\textwidth]{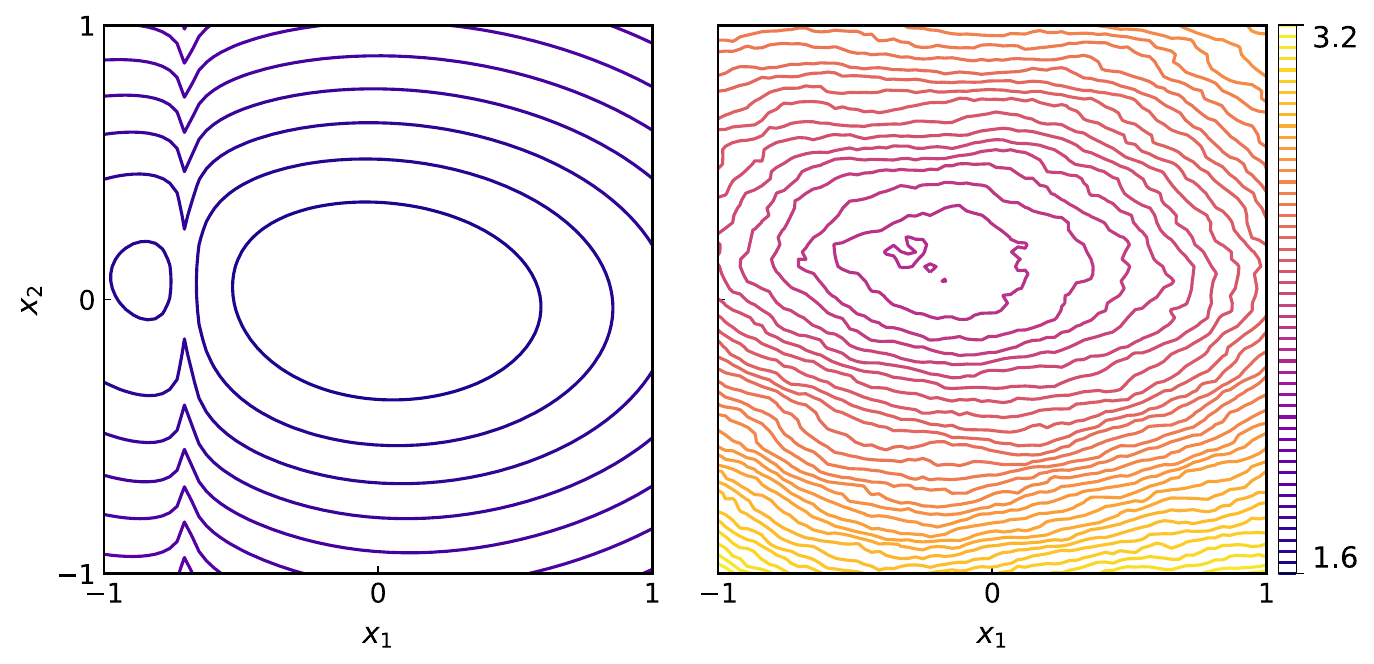}
         \caption{$R = 0.1$ noise}
     \end{subfigure}

     \vspace{10pt} % Adds some vertical spacing between the rows

     \begin{subfigure}[b]{0.49\textwidth}
         \centering
         \includegraphics[width=\textwidth]{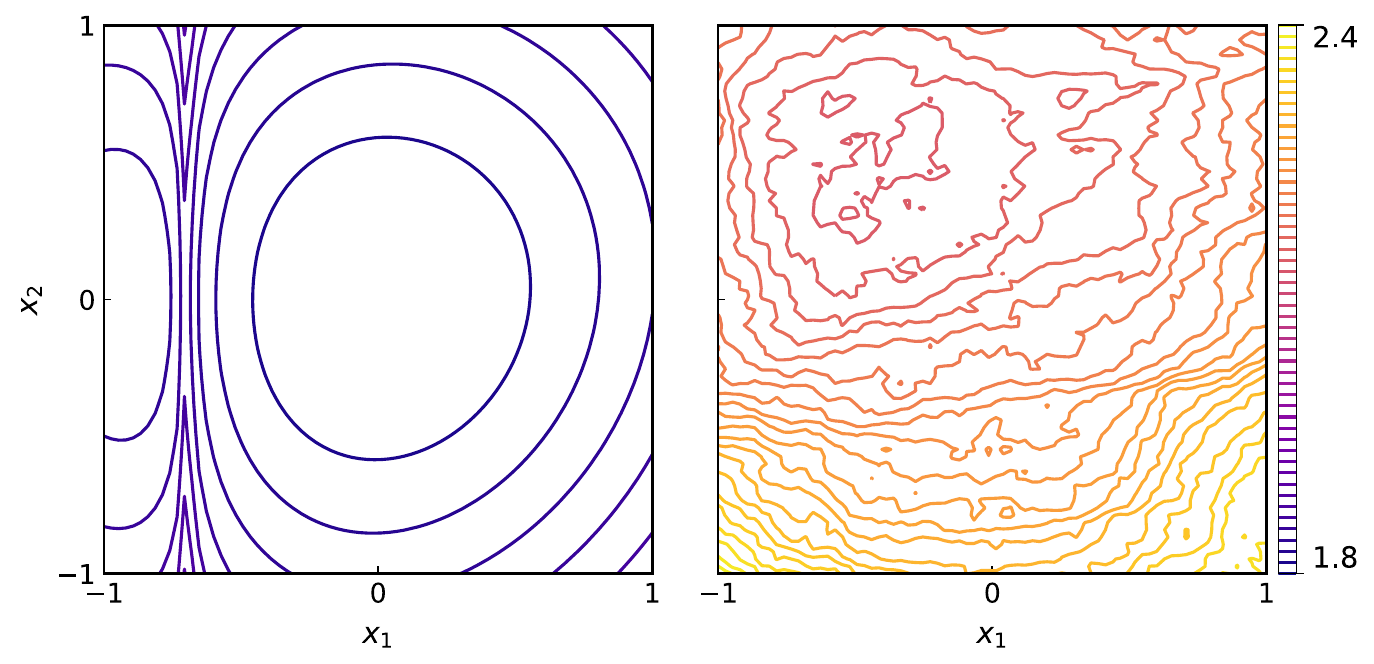}
         \caption{$R = 1$ noise}
     \end{subfigure}
     \hfill
     \begin{subfigure}[b]{0.49\textwidth}
         \centering
         \includegraphics[width=\textwidth]{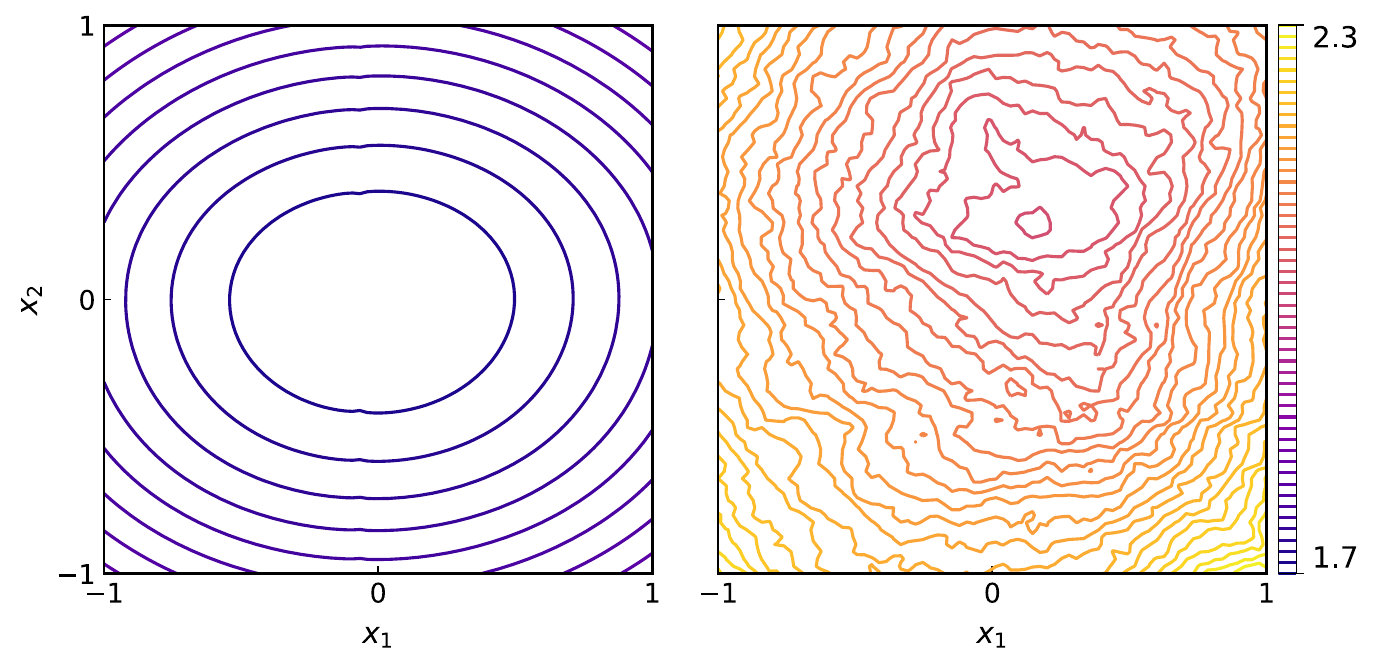}
         \caption{$R = 10$ noise}
     \end{subfigure}
     
     \caption{Initial loss landscape of two random orthogonal directions in the Hamiltonian variational subspace of $V$ defined in Eq.~(\ref{eq:variational_subspaces}) for the vanilla (left) architecture and the NDE architecture (right), for all four noise-to-unitary ratios $R = 0.01, 0.1,1, 10$. In all four cases, notice the heatmap of the contour lines indicating a barren landscape on the vanilla architecture, against a much richer loss-landscape for the NDE. This explains the performance gains of the NDE's robustness from Fig.~\ref{fig:NA_TFIM_robustness}, corresponding to the experiments in the top-left of Fig.~\ref{fig:NA_TFIM_robustness}}
     \label{fig:NA_TFIM_phs_H_loss_landscape}
\end{figure}

\begin{figure}
     \centering
     \begin{subfigure}[b]{0.49\textwidth}
         \centering
         \includegraphics[width=\textwidth]{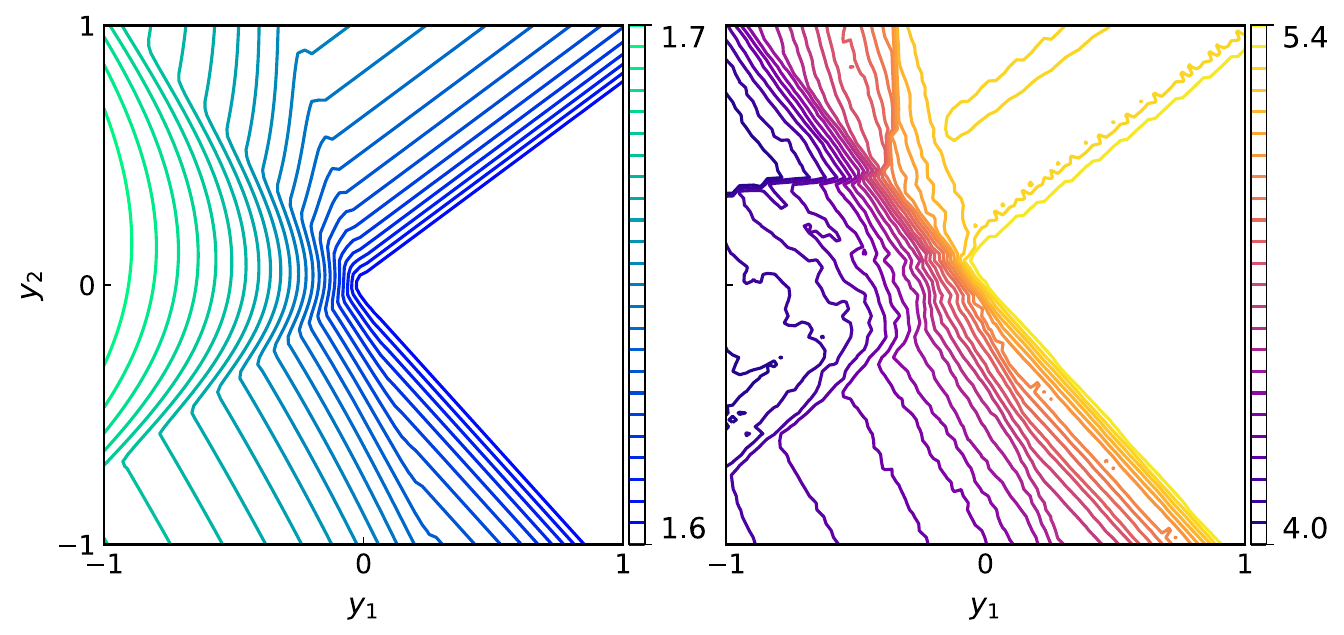}
         \caption{$R = 0.01$ noise} % (t = 10)
     \end{subfigure}
     \hfill
     \begin{subfigure}[b]{0.49\textwidth}
         \centering
         \includegraphics[width=\textwidth]{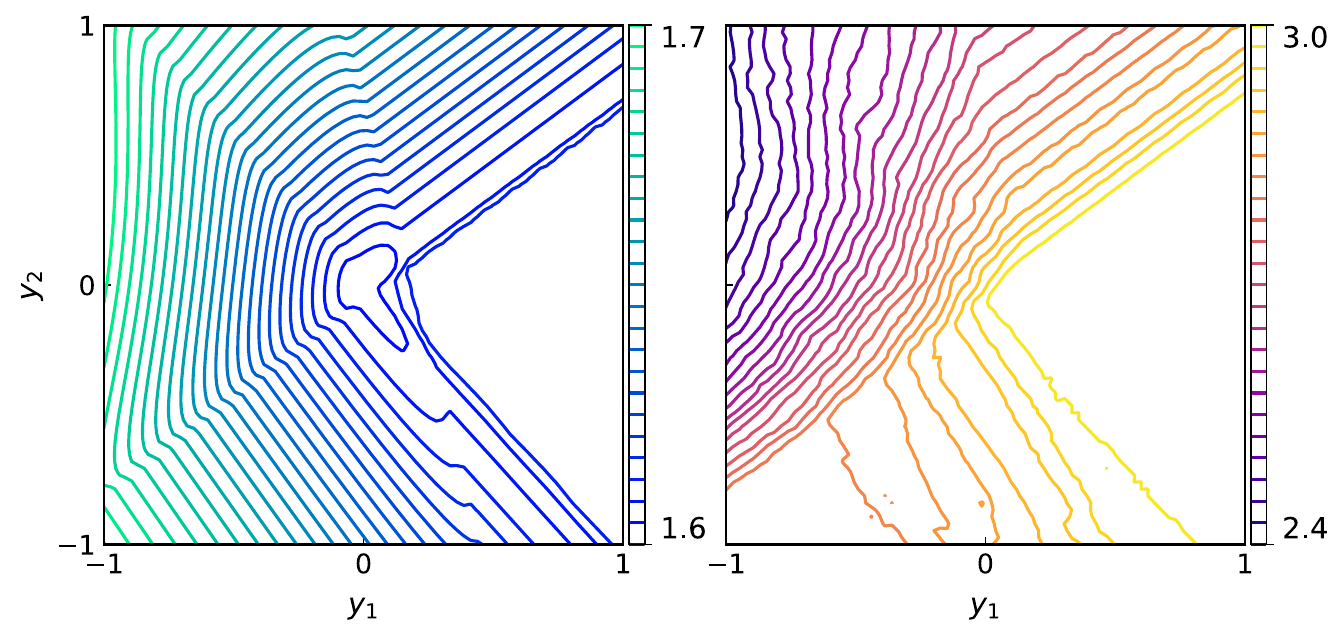}
         \caption{$R = 0.1$ noise}
     \end{subfigure}

     \vspace{10pt} % Adds some vertical spacing between the rows

     \begin{subfigure}[b]{0.49\textwidth}
         \centering
         \includegraphics[width=\textwidth]{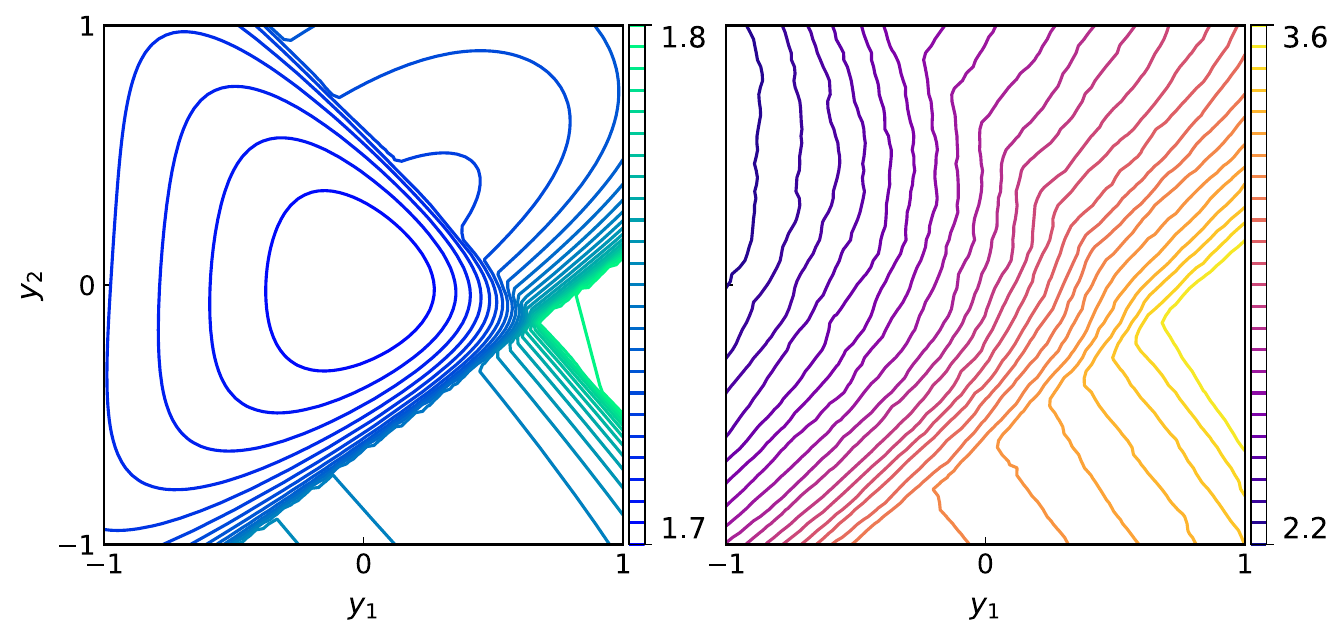}
         \caption{$R = 1$ noise}
         \label{subfig:NA_loss_landscape_phs_L}
     \end{subfigure}
     \hfill
     \begin{subfigure}[b]{0.49\textwidth}
         \centering
         \includegraphics[width=\textwidth]{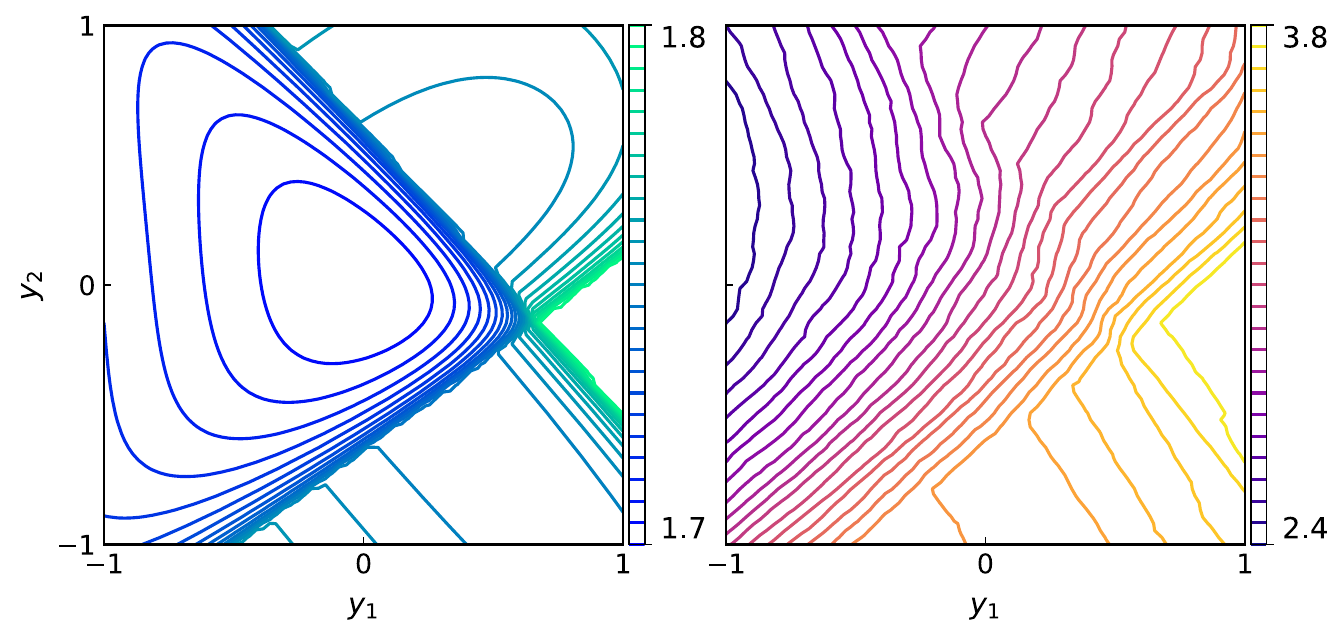}
         \caption{$R = 10$ noise}
     \end{subfigure}
     
     \caption{Initial loss landscape of two random orthogonal directions in the Lindbladian variational subspace of $V$ defined in Eq.~(\ref{eq:variational_subspaces}), for the vanilla (left) architecture and the NDE architecture (right), over all four noise-to-unitary ratios $R = 0.01, 0.1,1, 10$. In all four cases, notice the heatmap of the contour lines indicating a barren landscape on the vanilla architecture, against a much richer loss-landscape for the NDE. The legend for each heatmap shows the NDE having a landscape with heights that vary $\sim 14\times$ that of the vanilla model. This explains the performance gains of the NDE's robustness from Fig.~\ref{fig:NA_TFIM_robustness}, corresponding to the experiments in the bottom-left of Fig.~\ref{fig:NA_TFIM_robustness}}
     \label{fig:NA_TFIM_phs_L_loss_landscape}
\end{figure}

In Figs~\ref{fig:NA_TFIM_phs_H_loss_landscape}-\ref{fig:NA_TFIM_phs_L_loss_landscape}, we plot the loss landscape for our model upon initialisation for the Hamiltonian and dissipative parameters respectively in the case of phase noise. Each sub-figure represents one of the four orders of magnitude of noise considered in our work. Figs~\ref{fig:NA_TFIM_phs_H_loss_landscape}-\ref{fig:NA_TFIM_phs_L_loss_landscape} show two contrasting properties between the vanilla and NDE architecture. First, notice that in all cases, the vanilla loss-landscape is more barren for both the Hamiltonian and dissipative parts. Fig.~\ref{fig:NA_TFIM_phs_H_loss_landscape} plots contours with the same heat-map for both vanilla and NDE models, and indeed we see a much flatter landscape. This makes it more likely for the vanilla model to get stuck during optimization (rather than converge). Even if we renormalise the vanilla loss-landscape, as is done for dissipative part in Fig.~\ref{fig:NA_TFIM_phs_L_loss_landscape}, we see often the landscape has ridges within this variation of heights that is $\approx 14\times$ smaller than that of the NDE model. This shows us concretely why the NDE model was able to see the performance gains in Fig.~\ref{fig:NA_TFIM_robustness}.

Inspecting Figs~\ref{fig:NA_TFIM_phs_H_loss_landscape}-\ref{fig:NA_TFIM_phs_L_loss_landscape} further, we might wonder whether the NDE landscape is too rugged for effective learning. This challenges the notion that the NDEs are somehow smoothening the landscape and/or making it steeper. However, thus far our analysis has neglected the fact that the loss landscape of the Hamiltonian and dissipative parameters is \textit{dynamic} and changes between epochs. This is because by plotting the Hamiltonian and dissipative parameters, we are seeing a subset of the total variational space in both the vanilla and NDE architectures from Fig.~\ref{fig:architecures_all}. This follows from the fact that the total variational set for the vanilla model is $\theta = \{c_j, \gamma_{\alpha}\}$, and two random orthogonal directions in the Hamiltonian ($c_j$) or Lindbladian sets $(\gamma_{\alpha})$ do not have full support in the total variational space $\{c_j,\gamma_{\alpha}\}$. Likewise, in the NDE architecture, these remain subsets of an even larger variational set $\{\theta,\varphi\}$. During training, \textit{all} parameters in the variational set are able to change, thus the complement of two random orthogonal directions in $\{c_j\}$ or $\{\gamma_j\}$ changes and the loss landscape is not stationary. We will see this in more detail in Sec.~\ref{sec:XYZ_results} (see also, Fig.~\ref{fig:XYZ_loss_in_epochs}). 

For the combined noise case, we see performance degrading on both the vanilla and NDE model as explained in Sec.~\ref{sec:results_intro}.

% \tim{@ROMAN we should choose a different example. the one you chose already shows a neural landscape which is relatively smooth. I.e you did \ref{subfig:NA_loss_landscape_phs_L} but a better option is probably the (a) subfigure given that the NN "looks worse" for the reader.}

\begin{figure}
    \centering
    \includegraphics[width=\linewidth]{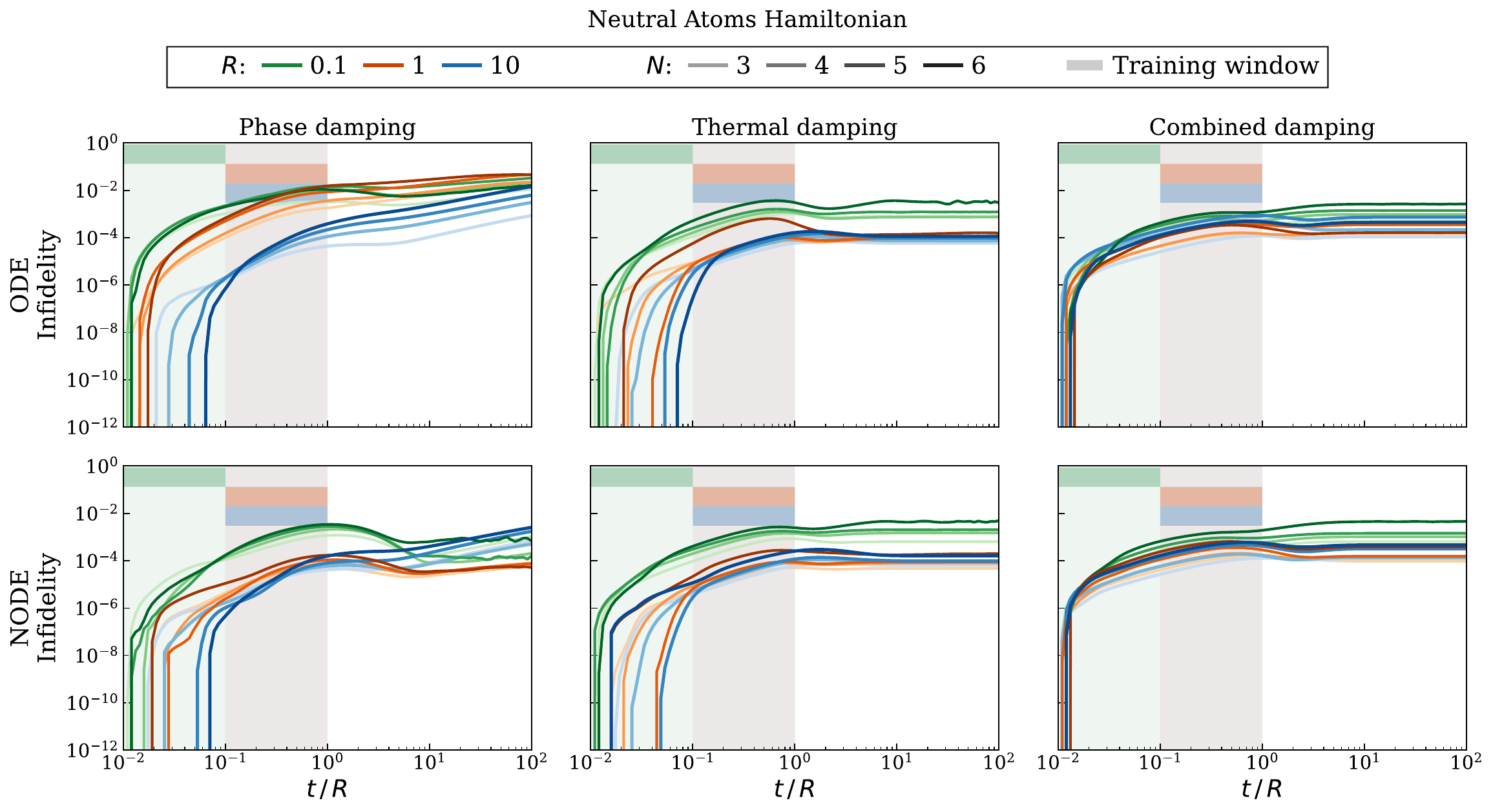}
    \caption{Infidelities against renormalised time for the vanilla (top row) and NDE (bottom row) architectures for the neutral atom TFIM with phase (left column), thermal (central) and combined (right) noise.  The shaded region corresponds to the training window, i.e. times over which snapshots were taken, and each system size $N = 3,4,5,6$ is plotted with increasing darker lines for $R = 0.1$ (green), $R = 1$ (red) and $R = 10$ (blue). Here, we renormalise by the noise-to-unitary ratio, $R$. We see NDE infidelities are shifted lower than the vanilla ones for phase (left) noise, with systems undergoing strong noise performing best as explained in the main text. Notice the vanilla and NDE models do not differ as much for the thermal (central) case, despite the increase in robustness shown in Fig.~\ref{fig:NA_TFIM_robustness}. This is due the unique steady-state effect explained in the main text.}
    \label{fig:NA_infidelities_all}
\end{figure}

Finally, let us turn our attention to the infidelity benchmarks, shown in Fig.~\ref{fig:NA_infidelities_all} for phase and thermal noise respectively. Fig.~\ref{fig:NA_infidelities_all} shows the NDE model improving the infidelity in both the transient (shaded) windows, as well as converging to a steady state which is closer to the true steady state. This can be read off directly by comparing left-to-right each shaded line colour and noting the NDE architecture has a lower infidelity in all cases, for all times simulated. Of note here is that the infidelity for $R = 0.01$ noise amplitude ratio remains below $0.1\%$ for the entire evolution window, which runs up to $10^3\times$ for all values of $N$, converging to a steady state with infidelity $\sim 10^{-4}$. This means our algorithm's learned action for $0.01\times$ phase noise can propagate up to $N=6$ qubit dissipative dynamics with fidelity $\mathcal{F} \geq 0.999$ for times up to $1000\times$ the training window. 

Meanwhile Fig.~\ref{fig:NA_infidelities_all} shows a comparable performance in infidelity between the vanilla and NDE model. This is despite the fact that Fig.~\ref{fig:NA_TFIM_robustness} showing the NDE model giving significant performance gains in terms of how robustly it can learn the values of the true parameters. This is because when the vanilla ODE model failed to converge, it is still in the neighbourhood of the true Hamiltonian/dissipative parameters, and the parameters it gets stuck at correspond to dynamics with the same fixed point as dynamics governed by the true parameters. As such, despite improving learnability with thermal noise, we do not see an obvious change in infidelity for this set of experiments. This highlights an important subtlety with the infidelity metric: the parameters can fail to converge to the correct parameters and this is not visible in infidelity whenever the converged parameters correspond to a system that shares one or more steady states with the true system. Indeed, this is why the system's infidelity tends to stabilise at a steady state which is close in Fidelity to the true steady-state. 

\subsection{Superconducting Hamiltonian}
\label{sec:SC_results}

\begin{figure}
    \centering
    \includegraphics[width=1\linewidth]{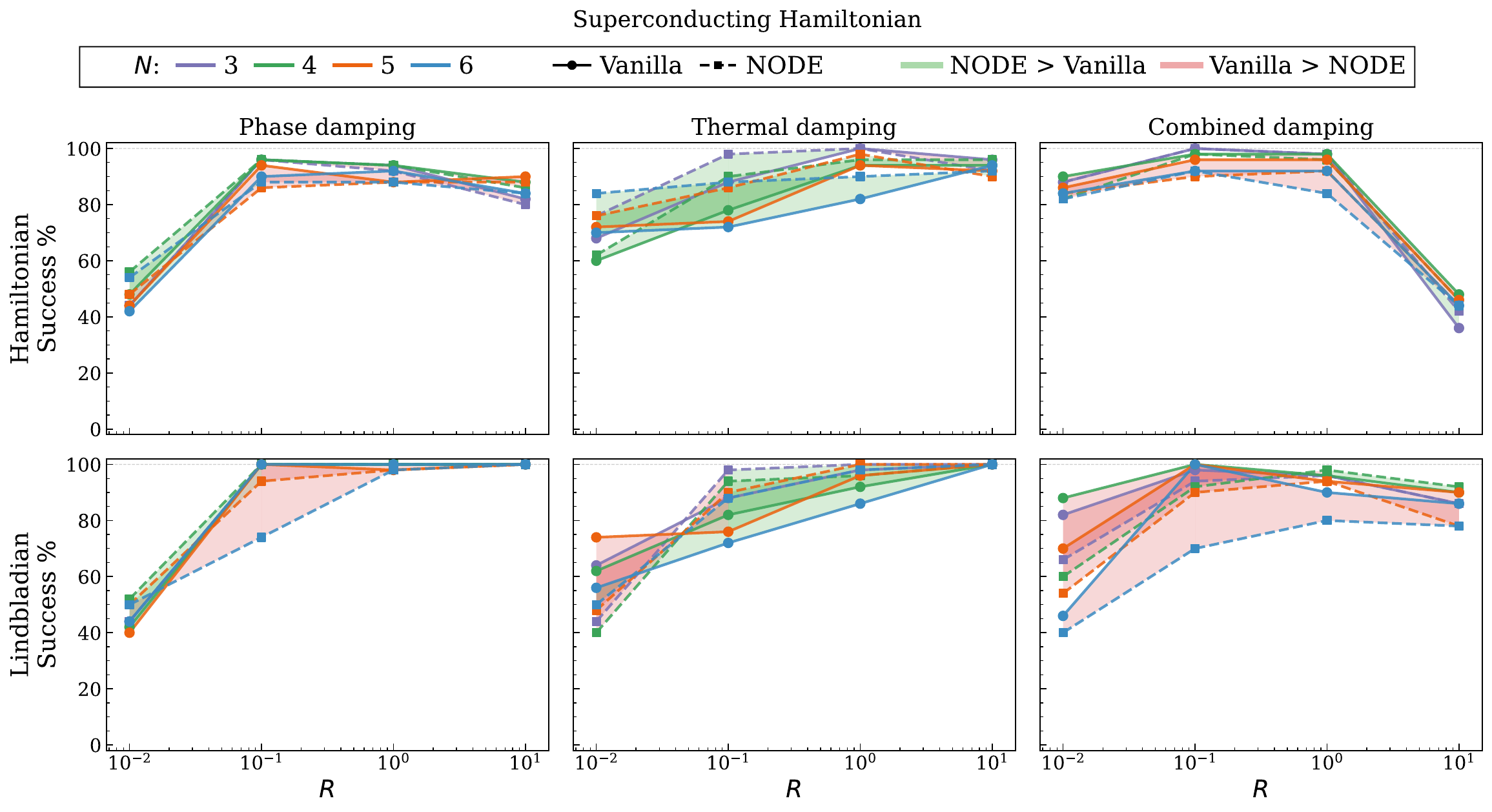}
    \caption{Success rates for a 1D superconducting Hamiltonian model with thermal, phase, and combined noise terms as a function of noise-to-unitary ratio $R$ over four orders of magnitude. The top row contains the superconducting model's Hamiltonian errors and the bottom row contains its Lindbladian errors. These are shown from left to right for (a) phase damping, (b) thermal damping and (c) combined damping. Shaded regions correspond to when the NDE term increases (red) or decreases (green) the fitting error. }
    \label{fig:SC_robustness}
\end{figure}

In contrast to neutral atoms, the superconducting model shows NDEs offer no performance gain with phase noise, sometimes even worsening the robustness, especially for $N = 6$. Fig.~\ref{fig:SC_robustness}'s left column shows the vanilla model performing better than the NDE model on both the Hamiltonian and dissipative parameters when the system is undergoing phase noise.  Indeed, when we analyse the commutant of the dissipative component of Eq.~\ref{eq:lindblad}, we find that for phase noise, the commutant is trivial, and there is a single unique steady state. This likely simplifies the learning task, rendering the standard Lindblad term sufficient and expressive enough to properly explore the variational landscape. Adding a neural correction term here is therefore training the neural variational parameters to \textit{overfit} to artifacts of the training dataset, rather than to the dynamics. 

On the other hand, the central column in Fig.~\ref{fig:SC_robustness} shows the NDE improving robustness in all cases of thermal noise except for the Lindbladian parameters at $N = 3$. Unlike phase noise, which commutes with the Superconducting Hamiltonian (and hence has a trivial commutant), thermal noise does not. This non-commutativity increases the complexity of the dynamics, and indeed there are several steady states in this setting \cite{Davies1974}. In this regime, the NDE ansatz becomes beneficial, with the improvements of the NDE ansatz increasing with $N$. 

Finally, the right hand column of Fig.~\ref{fig:SC_robustness} corresponds to both phase and thermal noise. It the NDE worsening the robustness marginally on the Hamiltonian parameters, and significantly for the Lindbladian ones. Despite this joint-dissipation having a component with a non-trivial commutant, the smoothing effect of the phase noise dominates, rendering the system solvable with the vanilla ODE alone. An further reason for this worsening is the NDE may be absorbing the noise dynamics itself rather than correcting the evolution based on fixed dissipators. To mitigate this, the regularization penalty on the NDE terms could need to be increased to force the model to prioritize the physical Lindblad terms over the NDE correction.

Let us now turn our attention to the infidelity benchmarks. Since the vanilla ODE architecture was robust on all noise scales $R = 0.1, 1, 10$, we see little to no difference between the vanilla and NDE models for both phase and thermal noise. We highlight here that like the neutral atoms experiments, when there is a failure to converge, the system still relaxes to the steady state of the true system despite. This is because when the steady state is unique. This indicates the fixed point of Eq.~\ref{eq:lindblad} is heuristically invariant to these small differences between true and learned parameters. Consequently, the variational parameters often get stuck in the neighbourhood of the true parameters, but do not converge because the dynamics are insensitive to the last updates needed to converge.

\begin{figure}
    \centering
    \includegraphics[width=\linewidth]{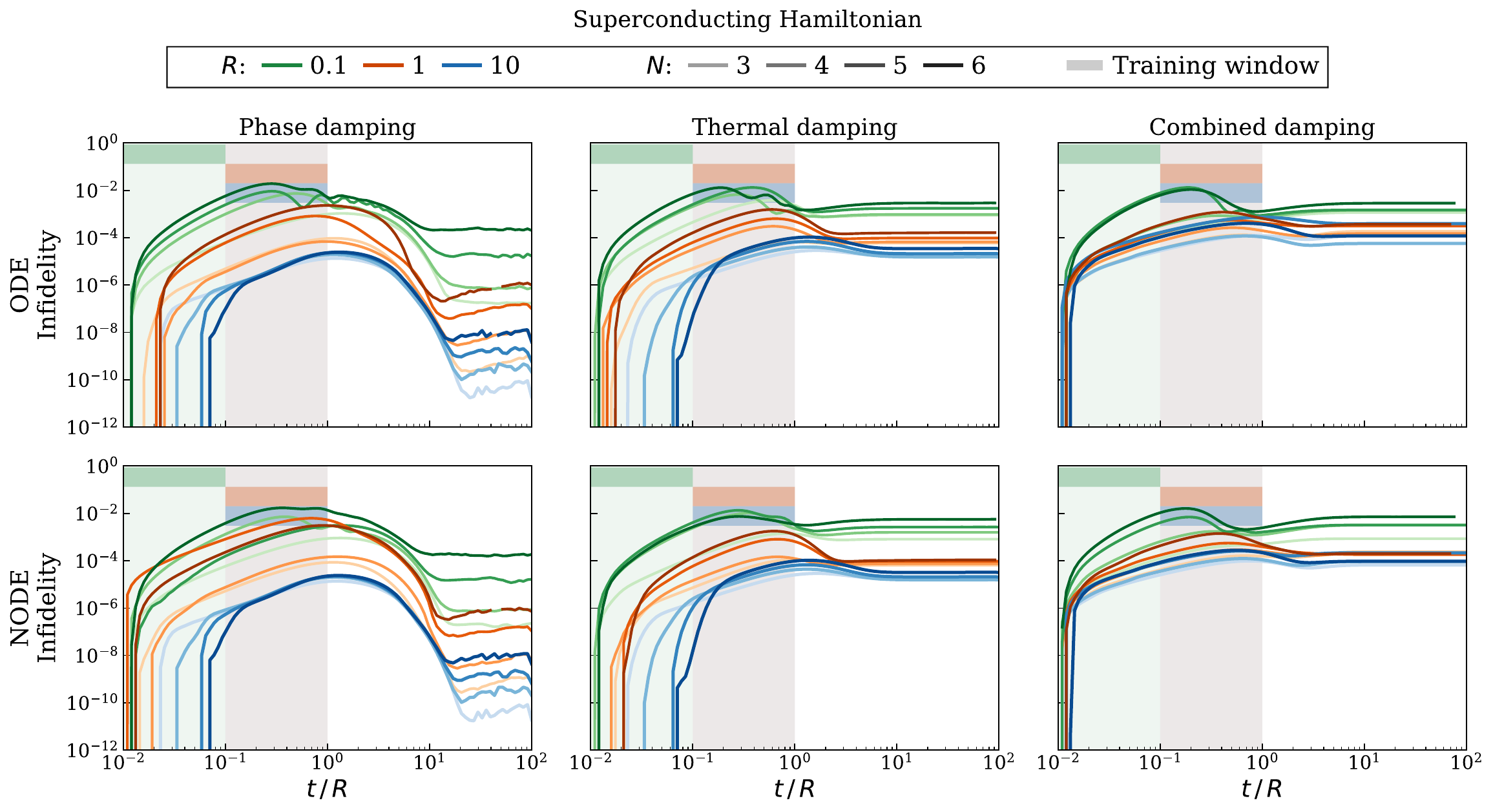}
    \caption{Infidelities against renormalised time for the vanilla (top row) and NDE (bottom row) architectures for the superconducting system with phase (left column), thermal (central) and combined (right) noise.  The shaded region corresponds to the training window, i.e. times over which snapshots were taken, and each system size $N = 3,4,5,6$ is plotted with increasing darker lines for $R = 0.1$ (green), $R = 1$ (red) and $R = 10$ (blue). Here, we renormalise by the noise-to-unitary ratio, $R$.}
    \label{fig:placeholder}
\end{figure}

\subsection{XYZ Heisenberg Model}
\label{sec:XYZ_results}

\begin{figure}[h!]
    \centering
    \includegraphics[width=1\linewidth]{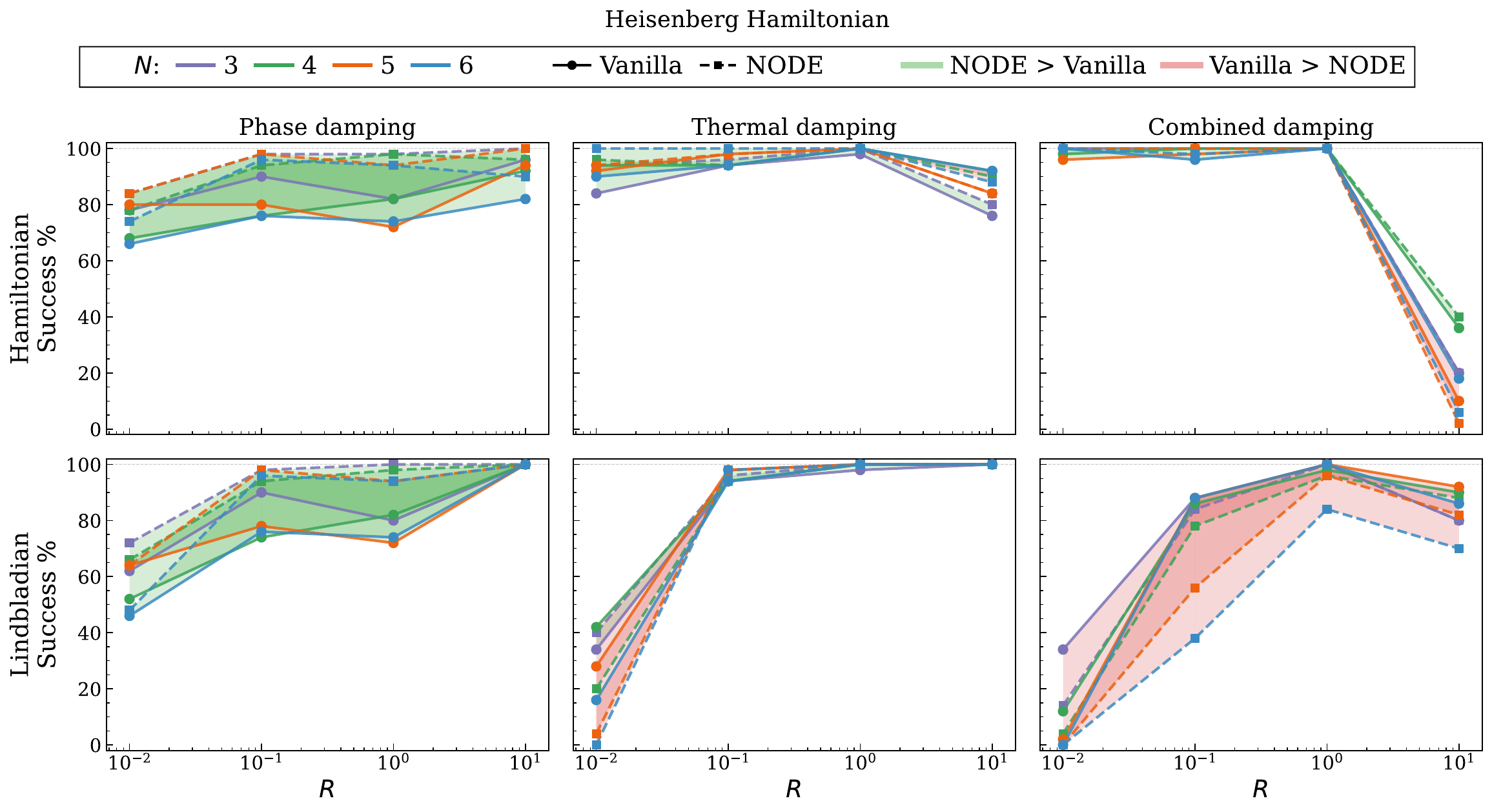}
    \caption{Success rates for the Heisenberg XYZ model with thermal, phase, and combined noise terms as a function of noise ratio $R$ over four orders of magnitude. The top row contains the XYZ model errors and the bottom row contains the noise model errors. These are shown from left to right for (a) phase damping, (b) thermal damping and (c) combined damping. Shaded regions correspond to when the NDE term increases (red) or decreases (green) the fitting error.}
    \label{fig:XYZ_robustness}
\end{figure}

Having now analysed the neutral atom and superconducting systems, it is clear that the type of Hamiltonian and noise affects both learnability, \textit{and} the ability of NDEs to improve trajectory-based-learning  prospects. Indeed this is most clear in the XYZ system. Here, phase noise is helped by NDEs as shown by Fig.~\ref{fig:XYZ_robustness}, where the left column shows the NDE improving robustness for all noise levels, on both the dissipative and unitary components. However, thermal noise is not helped by NDEs, and indeed they even worsened robustness for the dissipative parameters as seen by the central column of Fig.~\ref{fig:XYZ_robustness}. In line with the above two systems, we see a trend here on the complexity of dynamics and whether the NDE can help. On the XYZ model, trivial commutant of the thermal operator likely simplifies the learning task, rendering the standard Lindblad term sufficient until the signal is too weak to fit the dissipative parameters. Adding an NDE does not change the learning signal (i.e. training datasets) thus we see no gains by adding an NDE correction. On the other hand, the XYZ model with phase noise has a non-trivial commutant. As such, there are many-steady states, so having an NDE correction could help re-connect the different fixed points of the evolution. 

Consistent with the neutral atom and superconducting systems, we see that combined noise is also worsened by NDEs. This is again an over-parametrisation problem. Strong noise means the system thermalises quickly and there is a good signal for the noise terms that makes NDEs unnecessary, and they are representing dynamics that could have been attributed to the dissipative parameters were the NDE term not present. 

\begin{figure}[htbp]
    \centering
    \includegraphics[width=0.8\textwidth]{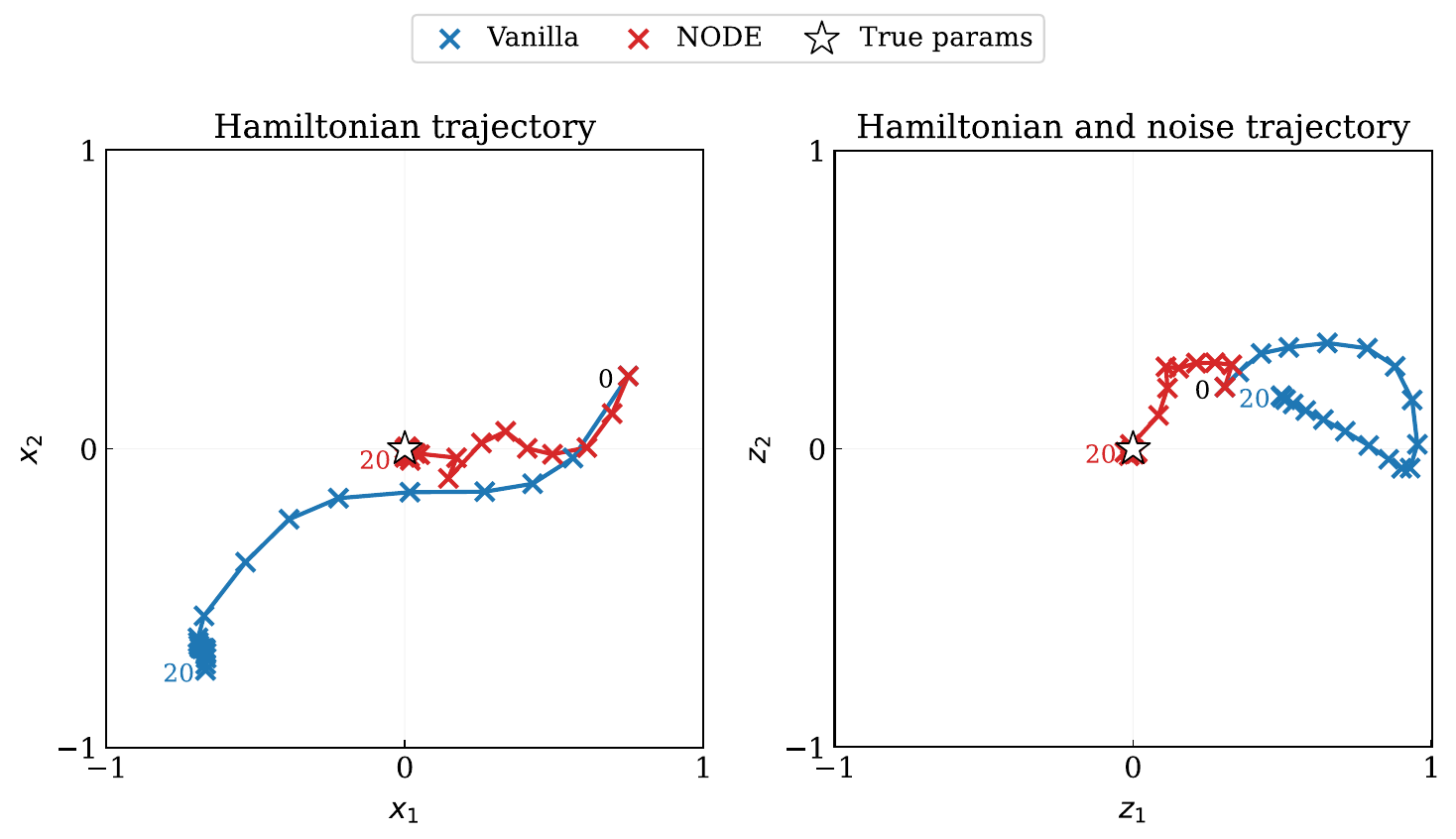}
    \caption{The trajectory of two random orthogonal directions in the Hamiltonian variational subspace of $V$ defined in Eq.~(\ref{eq:variational_subspaces}) (left) and the combined Hamiltonian and Lindbladian variational subspace of $V$ (right) for the XYZ Heisenberg model with $R=0.01$ noise-to-unitary ratio on $N = 3$ bodies. Comparing the vanilla (blue) and NDE architecture (red), we see the NDE architecture converging to the true values (set at the origin) in both cases. Whereas the vanilla architecture fails to converge in both cases, and its trajectory appears to get stuck, indicating the presence of local minima on those epochs of the vanilla model's training. Integer values on the plots refer to the start (0) and final (20) epochs.}
    \label{fig:param_traj_XYZ}
\end{figure}

\begin{figure}
    \centering
    \includegraphics[width=\linewidth]{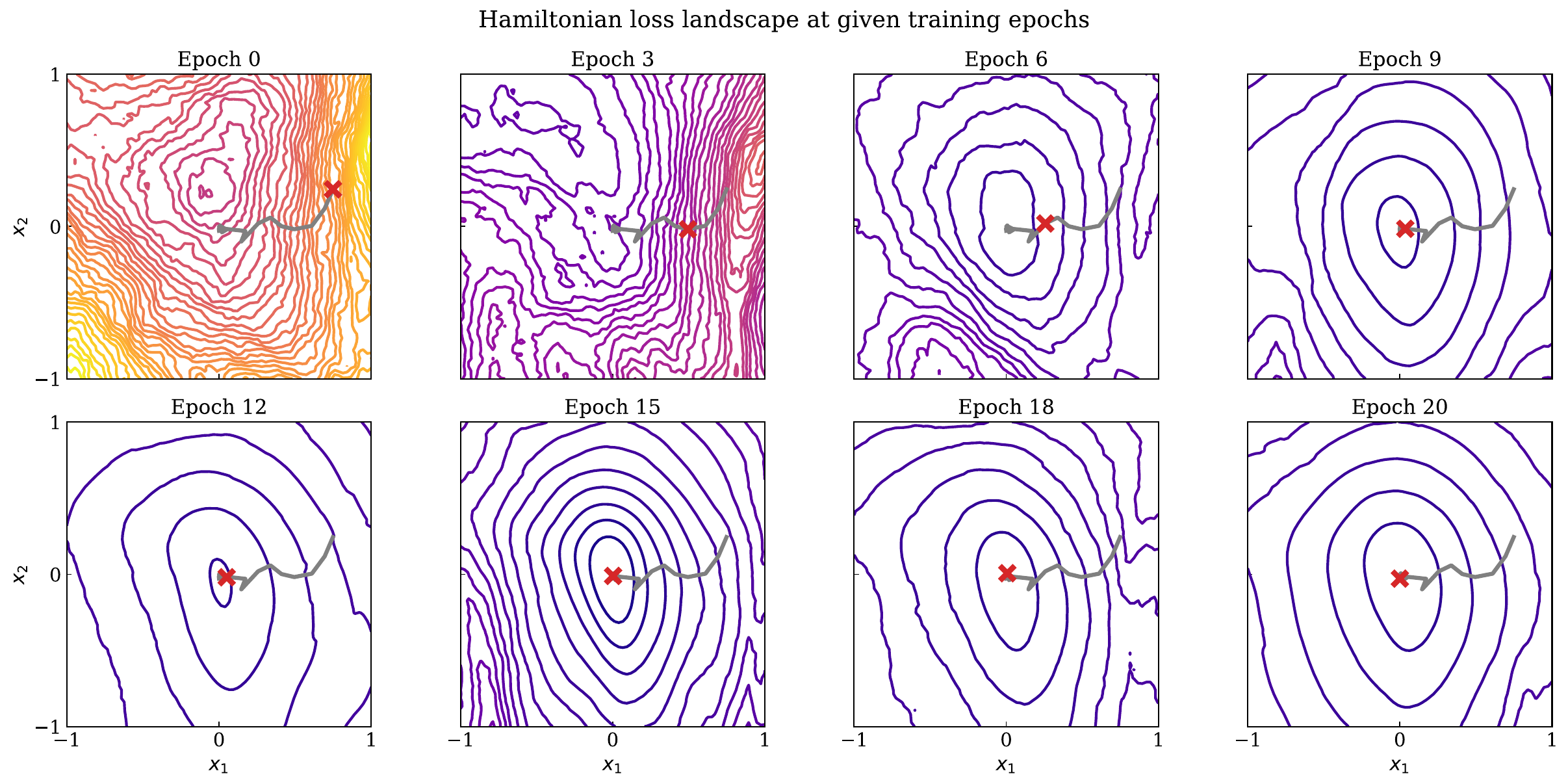}
    \caption{The loss-landscape of two random orthogonal directions in the Hamiltonian variational subspace of $V$ defined in Eq.~(\ref{eq:variational_subspaces}), for the XYZ Heisenberg Model at the start of epochs $0,3,6,9,12,15,18$ and $20$. The current parameter value shown by a red marker $\boldsymbol{\times}$. We see an initially rugged loss-landscape smoothing out as training proceeds until we arrive at a smooth landscape that is quasi-quadratic around the true value. The trajectory is shown by a grey line with the red $\times$ marking the current parameter values along the 20-epoch path.}
    \label{fig:XYZ_loss_in_epochs}
\end{figure}

Indeed, these effects are corroborated by the loss landscapes for the systems considered in Fig.~\ref{fig:XYZ_robustness}. Here, Fig.~\ref{fig:XYZ_loss_in_epochs} shows how the loss landscape evolves for one of the experiments. The landscape begins rugged, eventually smoothing into smooth landscape in which the parameters have converge to the true values at the origin. This means the overall trajectory for the NDE architecture converges to the true parameters, where the vanilla model fails. Plotting just the parameter values in epochs, as shown in Fig.~\ref{fig:param_traj_XYZ}, makes this immediately clear.

\begin{figure}
    \centering
    \includegraphics[width=\linewidth]{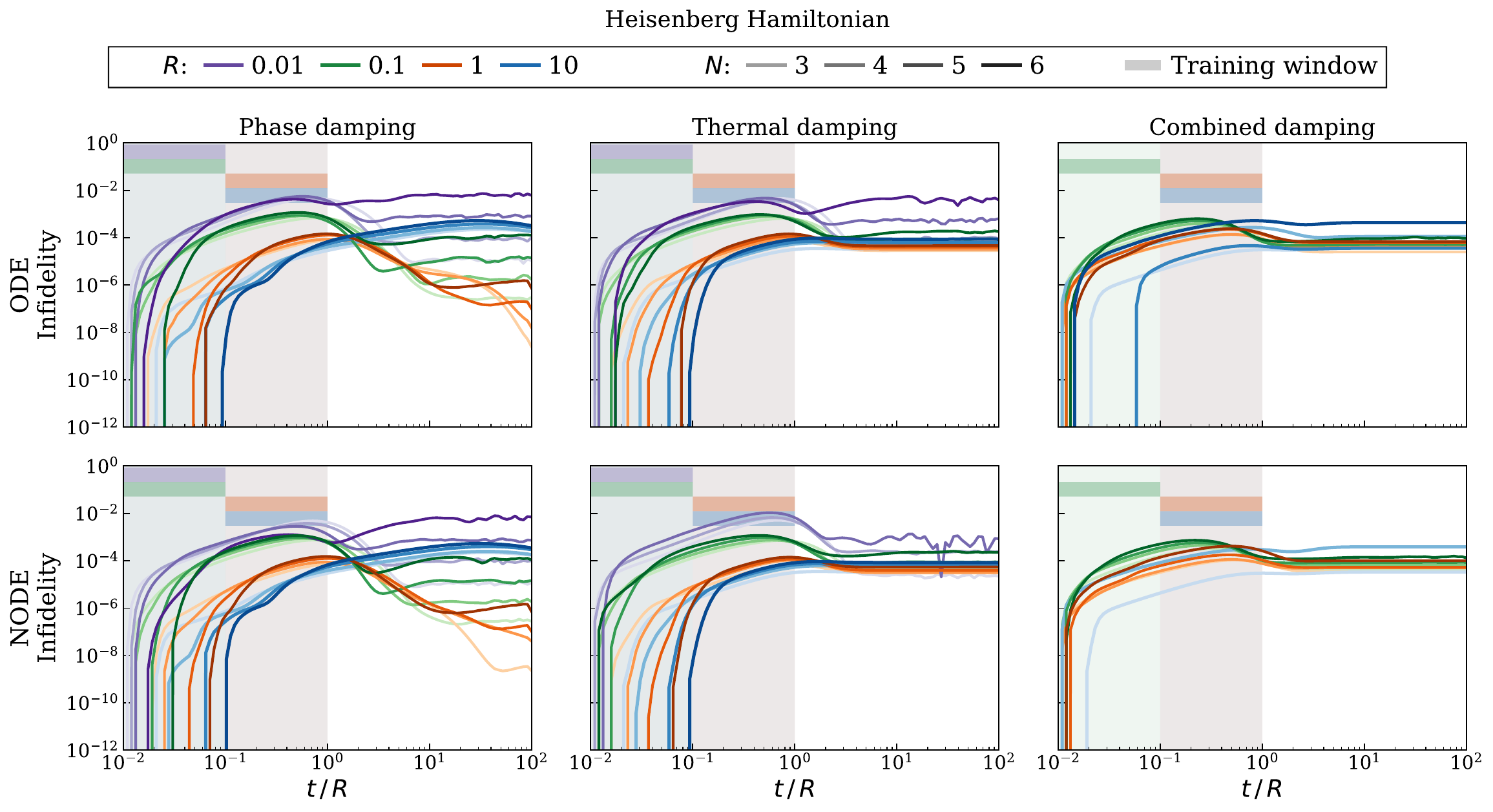}
    \caption{Infidelities against renormalised time for the vanilla (top row) and NDE (bottom row) architectures for the Heisenberg XYZ model with phase (left column), thermal (central) and combined (right) noise.  The shaded region corresponds to the training window, i.e. times over which snapshots were taken, and each system size $N = 3,4,5,6$ is plotted with increasing darker lines for $R = 0.1$ (green), $R = 1$ (red) and $R = 10$ (blue). Here, we renormalise by the noise-to-unitary ratio, $R$. There is little change in infidelity between these architectures, despite the robustness for phase noise improving in Fig.~\ref{fig:XYZ_robustness}. This is due again to the unique steady-state arising from this combined Hamiltonian and phase-noise model as explained in the main text - see also Sec.~\ref{sec:NA_results}.}
    \label{fig:XYZ_infidelity_phs}
\end{figure}

Next we again turn our attention to Figs.~\ref{fig:XYZ_infidelity_phs}, showing the infidelity in time for phase and thermal noise respectively. Per the neutral atom and superconducting case, we see a very similar performance on this benchmark between the vanilla and NDE correction. This is again because when the variational parameters are in the neighbourhood of the true parameters, the system will often converge to the same steady state as the true parameters. As such the ODE and NDE correction appear to be performing the same, despite the robustness experiments showing differences, see Secs.~\ref{sec:NA_results}~and~\ref{sec:SC_results} for a full discussion.

\subsection{PXP Hamiltonian}
\label{sec:PXP_results}
\begin{figure}
    \centering
    \includegraphics[width=1\linewidth]{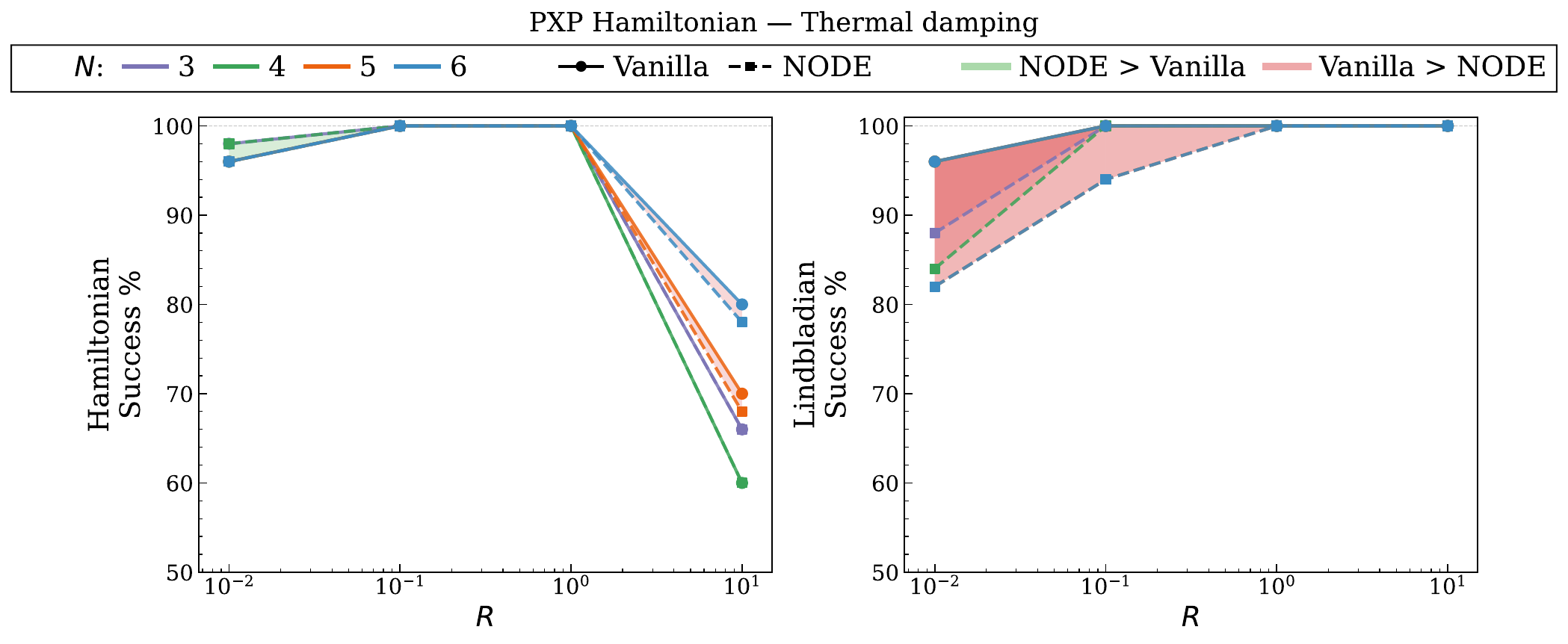}
    \caption{Success rates for Hamiltonian (left) and Lindbladian (right) parameters on the PXP model with thermal noise, as a function of noise ratio $R$ over four orders of magnitude. Solid lines represent vanilla architecture success rates and dashed lines represent NODE architecture success rates, with green (red) regions indicated the NODE architecture improving (diminishing) performance. Here we show $N = 3,4,5,6$.}
    \label{fig:PXP_robustness}
\end{figure}

\begin{figure}
    \centering
    \includegraphics[width=0.5\linewidth]{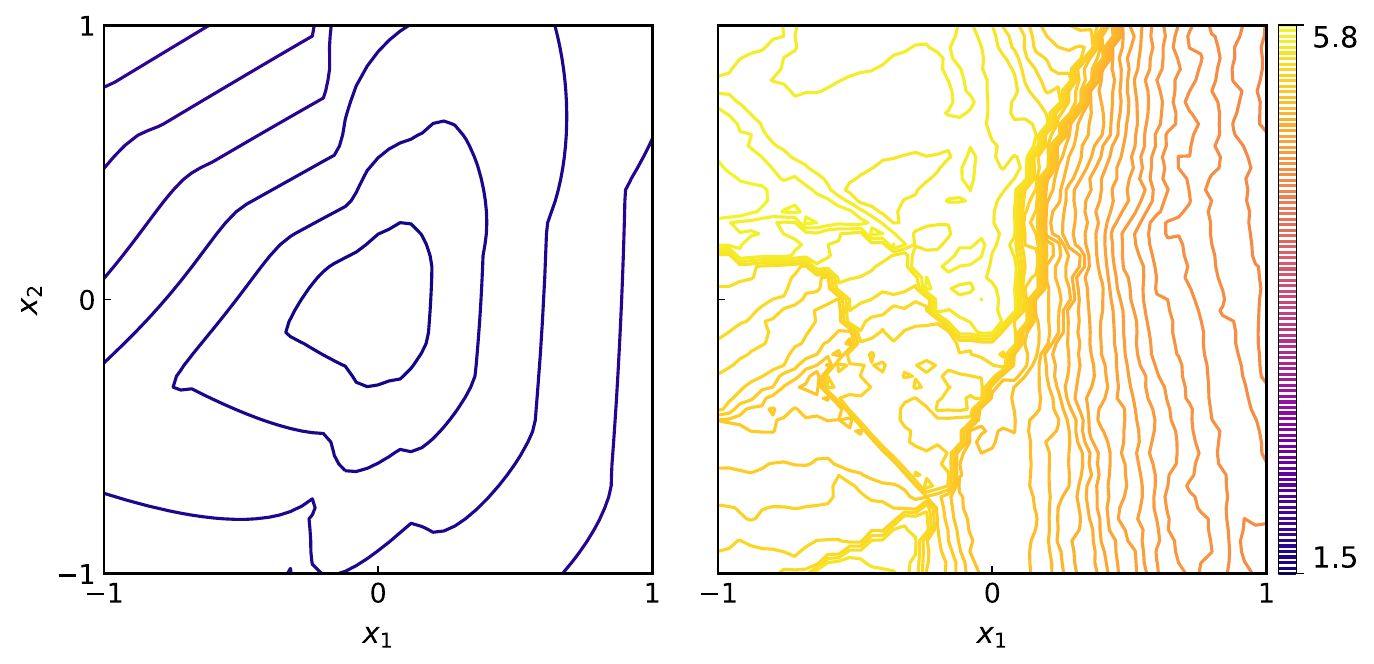}
    \caption{Initial loss landscape for two random orthogonal directions in the Lindbladian variational subspace of $V$ defined in Eq.~(\ref{eq:variational_subspaces}) for the PXP hamiltonian with thermal noise, with ratio $R = 0.01$. Here we plot the two random orthogonal variational directions on the dissipative component for the vanilla (left) architecture and the NDE architecture (right). We see the NDE's rugged landscape, contains many walls, is hindering convergence. This explains the drop in performance we see in the success rate figure for the PXP Hamiltonian with thermal noise, as explained in the main text.}
    \label{fig:PXP_loss_landscapes}
\end{figure}

\begin{figure}
    \centering
    \includegraphics[width=0.9\linewidth]{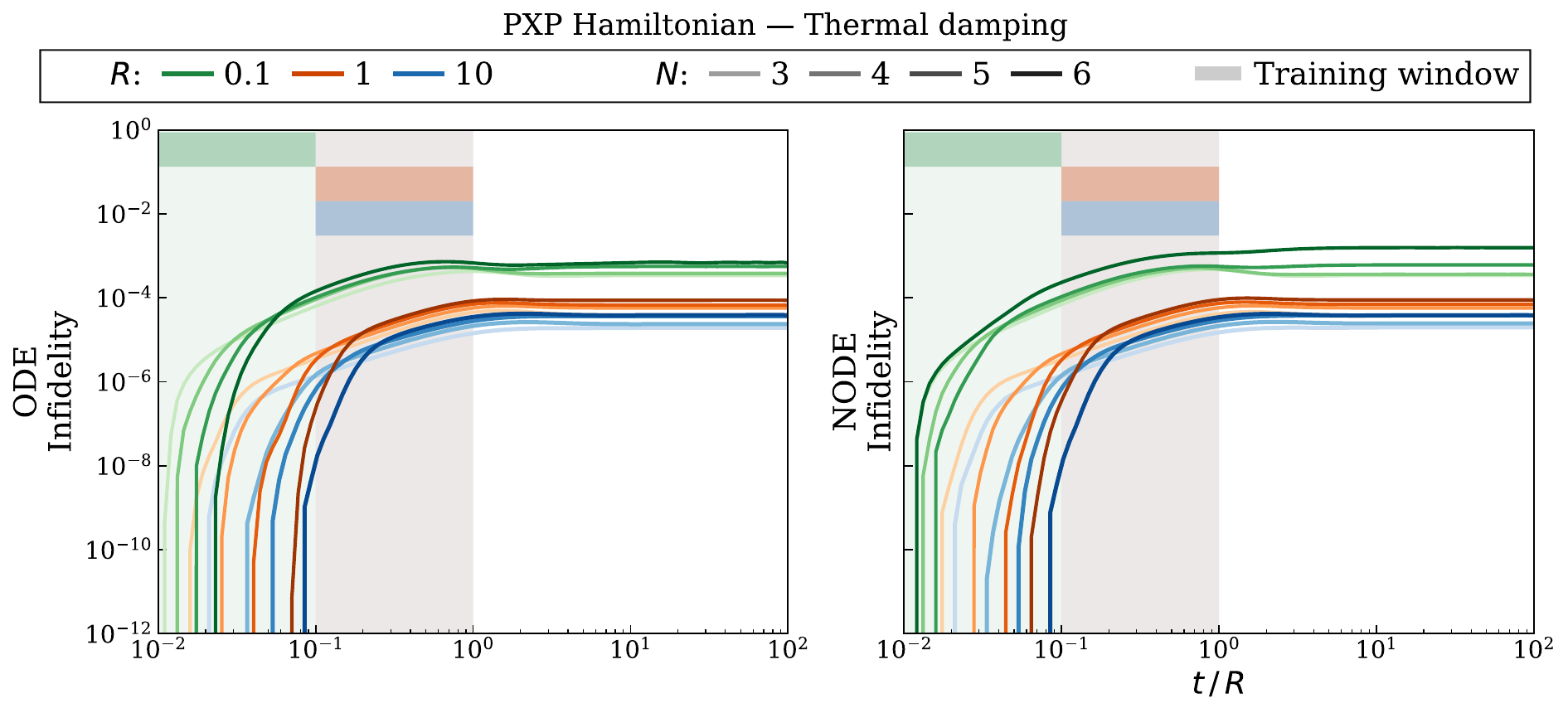}
    \caption{Infidelities against renormalised time for the vanilla (left) and NDE (right) architectures for the PXP  system undergoing thermal noise. Here, we renormalise by the noise-to-unitary ratio, $R$.  The shaded region corresponds to the training window, i.e. times over which snapshots were taken, and each system size $N = 3,4,5,6$ is plotted with increasing darker lines for $R = 0.1$ (green), $R = 1$ (red) and $R = 10$ (blue). There is little change in infidelity between these architectures, despite the robustness for phase noise improving in Fig.~\ref{fig:XYZ_robustness}. This is due again to the unique steady-state arising from this combined Hamiltonian and phase-noise model as explained in the main text - see also Sec.~\ref{sec:NA_results}}
    \label{fig:PXP_infidelities}
\end{figure}

On the PXP model with thermal noise, Fig.~\ref{fig:PXP_robustness} shows the robustness to be strong for both the vanilla and NDE models, with a slight degradation in solution quality for low noise on the dissipative parameters. As such, the vanilla model suffices on this system. The degradation of robustness of the NDE model's Lindbladian parameters is again visible in the loss landscapes. Fig.~\ref{fig:PXP_loss_landscapes} shows the initial loss landscapes for the vanilla and NDE model, with a considerably more rugged loss-landscape. This rugged landscape persists in the training epochs, and as such the parameters fail to converge. Finally, we show the infidelity benchmarks for this system in Fig.~\ref{fig:PXP_infidelities}, which show the learned parameters rapidly approaching a steady state. Of note here is that for the PXP system, the model's infidelity does not peak in the transient times of evolution. This is unlike the other three models tested, and is indicative of the PXP Hamiltonian with thermal noise having a strongly-attractive fixed point. This also sits in strong contrast with the PXP Hamiltonian alone (i.e. in the unitary case), in which it is weakly non-ergodic \cite{serbyn2021quantum}.

\section{Conclusion}
\label{sec:conclusion}

In this work we developed a trajectory-based method for Lindbladian learning from random local Pauli measurement snapshots collected at multiple \emph{transient} times. By fitting a maximum-likelihood objective to measurement outcomes, our approach avoids full process tomography while returning an interpretable Lindbladian whose Hamiltonian terms and jump-operator structure are fixed by a physically motivated model.

We adapted neural differential equations to open-system identification in a way that preserves interpretability: a neural correction aids optimization during an initial augmented training stage, after which a curriculum switches off the neural term and refines only the physical parameters. The final model therefore contains only Hamiltonian couplings and dissipative rates, with no residual neural component. Hence our algorithm returns a GKSL form that generates a valid CPTP map.

Across four Hamiltonian families (neutral-atom, superconducting, Heisenberg XYZ, and PXP), phase and thermal noise spanning four orders of magnitude, and system sizes up to $N=6$ qubits, we find a consistent regime structure. When coherent and dissipative contributions induce rugged, highly non-convex loss landscapes, the neural augmentation substantially increases the probability of successful recovery of both Hamiltonian and dissipative parameters. This was most prominent in settings where the coherent and dissipative components of the GKSL form are non-commuting. Conversely, when the physics-only landscape is already well-conditioned, physics-only training typically suffices and neural augmentation can degrade performance via overfitting. These observations motivate a practical guideline: attempt physics-only variational fitting first, and deploy neural augmentation only when robustness is insufficient.

Finally, we emphasize that open-system evaluation requires care. Low state infidelity can persist even when parameters are not reliably recovered, particularly when late-time behavior is dominated by steady-state structure. For this reason we advocate parameter-recovery robustness as the primary figure of merit in synthetic studies where ground truth is known, complemented by extrapolation tests beyond the training window. This sits in contrast to \cite{heightman2024solving} in which the power-law scaling of infidelity in unitary dynamics offered a proxy for the uncertainty in out-of-training times. 

The main limitation of our present implementation is how the curse of dimensionality affects scaling. Explicit density-matrix simulation grows exponentially with system size, meaning for $N=6$ our Hilbert space dimension is already $4^6 = 4096$. Indeed, the ultimate aim of Hamiltonian and Lindbladian learning is to characterize quantum devices directly from measurement data, without requiring such classical simulation in the full system's Hilbert space. A natural next step toward this goal is to use the present approach as a subroutine within locality-based or patch-learning schemes that learn overlapping reduced generators and stitch them into larger-scale models \cite{franceschetto2025hamiltonian, becker2021quantum}. The curriculum strategy also suggests a route toward settings where the operator structure is partially or fully unknown: the physical model can be progressively enriched, with the residual magnitude of the neural correction serving as a diagnostic of model misspecification. Indeed an interesting future research direction lies in using the relative contribution of the neural correction term as a proxy for expressivity of the interpretable ansatz. When the ansatz is sufficiently expressive to capture the full generator, the NODE component should converge to $0$ contribution with our curriculum. When this fails to be the case, it could be indicative of the ansatz lacking expressivity, and further terms should to be progressively added and tuned. This would introduce a different measure of learning-hardness based on the minimal Choi rank of the true dynamics; a higher-rank channel needs more Kraus operators and thus the ansatz' Choi rank could be steadily increased until it captures the dynamics and matches the Choi rank of the true channel. Finally, our analysis assumes time-homogeneous Markovian dynamics, known initial preparations, and ideal measurements; extending to time-dependent or non-Markovian generators and incorporating SPAM error models are also important directions for future work and experimental deployment. These considerations, while out of scope for this contribution, are a promising future research direction. 

As the second paper in a series of works (see also \cite{heightman2024solving}) systematically investigating NDEs in quantum system characterisation, our next works in this series seeks to address these scaling limitations, as well as the potential for this framework to be applied in black-box settings.

\bibliographystyle{unsrtnat}
\bibliography{references}
\end{document}

%% file: Figures/architectures/vanilla.tikz
\begin{tikzpicture}[
    every edge/.style = {draw,->}
  ]
  \node (start) at (0, 1.6) [draw, minimum width=3cm] {$\rho_0$};
  \node (state) at (0, 0) [draw, minimum width=3cm] {$\rho$};
  \node (h) at (3, 0) [draw] {$\mathcal{L}_{\theta}(\cdot)$};
  \node (times) at (2, -1.8) [draw, circle] {$\mathcal{L}_{\theta}(\rho)$};
  \node (end) at (2, -3.3) [] {};
  \draw (state) edge[out=270, in=90, looseness=1] (times);
  \draw (h) edge[out=270, in=90, looseness=1] (times);
  \draw (times) edge[] (end);
  \draw (start) edge[] (state);
  \node (ode) at (1.1, -0.9) [draw, minimum width=6cm, minimum height=3.5cm, label={[shift={(-2,-3.5)}]ODEint}] {};
\end{tikzpicture}

%% file: Figures/architectures/NDE.tikz
\begin{tikzpicture}[
    every edge/.style = {draw,->}
  ]
  \node (start) at (0, 1.6) [draw, minimum width=3cm] {$\rho_0$};
  \node (state) at (0, 0) [draw, minimum width=3cm] {$\rho$};
  \node (h) at (3, 0) [draw] {$\mathcal{L}_{\theta}(\cdot)$};
  \node (times) at (3, -2) [draw, circle] {$\mathcal{L}_{\theta}(\rho)$};
  \node (mlp) at (-1, -2) [draw, minimum height=1cm, minimum width=3cm] {$\text{NN}_{\varphi}(\rho)$};
  \node (add) at (2, -4) [draw, circle] {$+$};
  \node (end) at (2, -5.5) [] {};
  \draw (state) edge[out=270, in=90, looseness=1] (times);
  \draw (h) edge[] (times);
  \draw (state) edge[out=270, in=90, looseness=1] (mlp);
  \draw (times) edge[out=270, in=90, looseness=1] (add);
  \draw (mlp) edge[out=270, in=90, looseness=1] (add);
  \draw (add) edge[] (end);
  \draw (start) edge[] (state);
  \node (ode) at (0.6, -2) [draw, minimum width=7cm, minimum height=5.5cm, label={[shift={(-2.6,-5.5)}]ODEint}] {};
\end{tikzpicture}